\documentclass[12pt,a4paper]{article}
\usepackage{macros}

 \def\a{\alpha}

\preprint{}
\title{Notes on Quasinormal Modes of charged de Sitter Blackholes from Quiver Gauge Theories}

\author[a]{Pujun Liu}
\author[a,b]{Rui-Dong Zhu}
\affiliation[a]{School of Physical Science and Technology, Soochow University, 333 Ganjiang Road, 215006 Suzhou, China
}
\affiliation[b]{Institute for Advanced Study, Soochow University, 333 Ganjiang Road, 215006 Suzhou, China}
\emailAdd{pujunliu@hotmail.com}\emailAdd{rdzhu@suda.edu.cn}

\abstract{We give the connection formulae for ordinary differential equations with 5 and 6 (and in principle can be generalized to more) regular singularities from the data of instanton partition functions of quiver gauge theories. We check the consistency of these connection formulae by numerically computing the quasinormal modes (QNMs) of Reissner-Nordström de Sitter (RN-dS) blackhole. Analytic expressions are obtained for all the families of QNMs, including the photon-sphere modes, dS modes, and near-extremal modes. We also argue that a similar method can be applied to the dS-Kerr-Newman blackhole.}

\begin{document}

\allowdisplaybreaks

\maketitle

\section{Introduction}
Quasinormal modes (QNMs) give an important class of physical quantities that characterize the decaying feature of fields in blackhole background or the perturbations of blackhole geometry itself \cite{vishveshwara1970scattering,Kokkotas:1999bd,Berti:2009kk,Konoplya:2011qq,Horowitz:1999jd}. They play a key role in the study of the ring-down phase of gravitational wave merger events, and are also essential in the argument for the strong cosmic censorship problem in de Sitter blackholes \cite{Cardoso:2017soq}. QNMs are traditionally studied in various numerical methods, including Leaver's continued fraction method \cite{Leaver:1985ax,Leaver:1990zz,Nollert:1993zz}, WKB method \cite{Konoplya:2003ii}, isomonodromy method (refer to \cite{Cavalcante:2023rdy} for a review), Bender-Wu method \cite{Hatsuda:2019eoj,Eniceicu:2019npi} and the discretization of differential equation \cite{Gundlach:1993tp,Jansen:2017oag}. These numerical methods are usually technical and requires computations up to high-order. The discovery of the connection between (exactly solvable) supersymmetry gauge theories (or equivalently the Seiberg-Witten theory) in the so-called Nekrasov-Shatashvili limit and the perturbation equations in blackhole geometry  \cite{Aminov:2020yma}, however, opened up a new gate towards systematic and analytic study of the QNMs and other physical observables during the gravitational wave release. Since then intensive studies have been done especially focusing on the Schwarzschild and Kerr blackholes \cite{Hatsuda:2020sbn,Hatsuda:2020iql,Bonelli:2021uvf,daCunha:2021jkm,Bianchi:2021mft,Nakajima:2021yfz,Hatsuda:2021gtn,Fioravanti:2021dce,Bonelli:2022ten,Bianchi:2022wku,Dodelson:2022yvn,Consoli:2022eey,Imaizumi:2022qbi,Fioravanti:2022bqf,Lisovyy:2022flm,Bhatta:2022wga,daCunha:2022ewy,Gregori:2022xks,Imaizumi:2022dgj,Bianchi:2022qph,Fioravanti:2023zgi,Cano:2023tmv,Bianchi:2023rlt,Fucito:2023plp,Giusto:2023awo,Aminov:2023jve,Kwon:2023ghu,Hatsuda:2023geo,Bhatta:2023qcl,He:2023wcs,Lei:2023mqx,BarraganAmado:2023wxt,Fucito:2023afe,Bautista:2023sdf,DiRusso:2024hmd,Ge:2024jdx,Zhao:2024rrw,Cipriani:2024ygw,Arnaudo:2024rhv,Ishigaki:2024pfv,Bianchi:2024vmi,Bianchi:2024mlq,Jia:2024zes,Fucito:2024wlg,Aminov:2024aan,Aminov:2024mul,BarraganAmado:2024tfu,Matone:2024ytm,Ren:2024ifh}. 

One of the clearest explanations is provided in \cite{Bonelli:2022ten} on how the gauge-theory data help to solve ordinary differential equations (ODEs) with singularities, which correspond to the perturbation equations we usually see in blackhole geometries. In that work, it was shown based on the Alday-Gaiotto-Tachikwa (AGT) relation \cite{Alday:2009aq} that the instanton partition function of gauge theory gives the connection formula of the ODE, i.e. how the solutions expanded around one singularity are written as linear combinations of the solutions around other singularities. It is then straightforward to require some of the coefficients to be zero to satisfy the boundary conditions of the QNMs to solve them. More interestingly, in the previous study of accelerating blackhole (charged C-metric) \cite{Lei:2023mqx}, part of the QNMs are identified as the poles of $\Gamma$-functions in the coefficient of connection formula. If this argument is applicable to all the QNMs, it will provide not only an efficient numerical approach to QNMs, but will also allow one to systematically write down the analytic expression of QNMs in a much simpler way in terms of a quantum number $n=0,1,2,\dots\in \mathbb{N}$ corresponding to the $n$-th pole of the $\Gamma$-function, $\Gamma^{-1}(-n)=0$. In this article, we focus on the Reissner-Nordstr\"om de Sitter (RN-dS) blackhole, which has a rich mathematical structure as the playground for testing the numerical convergence of the connection formulae and also the analytic approach. In the asymptotically flat RN blackhole, analytic expressions for the near-extremal modes were found in \cite{Kim:2012mh} by utilizing the hidden SL(2,$\mathbb{R}$) symmetry of the geometry. It was further conjectured from the numerical study in \cite{Cardoso:2017soq} for the RN-dS blackhole that the leading-order behaviors of the near-extremal modes stay the same, and another family of QNMs, the dS modes are given the same value at the leading-order as those in pure dS geometry. In this article, we first numerically check the effectiveness of the connection formula obtained for ODEs with 5 and 6 regular singularities, and then show that one can obtain analytic expressions for not only the dS modes and near-extremal modes but also for the photon-sphere (PS) modes. The leading-order behavior of the former two families matches perfectly with those conjectured in the literature. 

The article is organized as follows. In section \ref{s:conn}, we give the connection formulae for ODEs with 5 and 6 regular singularities by generalizing the argument in \cite{Bonelli:2022ten}. Numerical results for scalar and non-scalar QNMs are respectively shown in section \ref{s:5pt-RNdS} and \ref{s:non-s}. We will present in detail how to derive the analytic results for scalar QNMs in section \ref{s:ana}, and it will be extended straightforwardly in section \ref{s:non-s} to non-scalar PS modes. 

\section{Connection formulae for ODE with regular singularities}\label{s:conn}

In this section, we review a systematic way proposed in \cite{Bonelli:2022ten} to write down the connection formulae for ODEs with only regular singularities, and write down the explicit formulae for 5 and 6 regular singularities. 

We first notice that the ODE with $n$ regular singularities can be obtained in the semiclassical limit $c\to \infty$ of the Belavin-Polyakov-Zamolodchikov (BPZ) equation satisfied by $n$-pt correlation function plus one degenerate field insertion in 2d CFT, 
\begin{equation}
    \lt(b^{-1}\frac{\partial^2}{\partial z^2}+\sum_{i=1}^n\frac{\partial_{t_i}}{z-t_i}+\frac{\Delta_i}{(z-t_i)^2}\rt)\langle V_{2,1}(z)\prod_{i=1}^nV_{i}(t_i)\rangle=0,
\end{equation}
where $\Delta_i$ stands for the conformal weight of the $i$-th primary field, the central charge of the 2d CFT is parameterized as 
\begin{equation}
    c=1+6\cQ^2,\quad \cQ=b+b^{-1},
\end{equation}
and we will take the semiclassical limit by sending $b\to 0$ with $b^2\Delta_i$ kept finite. One can fix three points by putting say $t_1=0$, $t_4=1$ and $t_5=\infty$ without losing generality, and then the BPZ equation for $n=5$ becomes 
\begin{align}
    \lt(b^{-2}\partial_z^2+\frac{\Delta_4}{(z-1)^2}-\frac{\sum_{i=1}^4\Delta_i+\Delta_{2,1}+z\partial_z-\Delta_5+t\partial_t+q\partial_q}{z(z-1)}+\frac{\Delta_3}{(z-t)^2}+\frac{\Delta_2}{(z-q)^2}\rt.\cr
    \lt.+\frac{t\partial_t}{z(z-t)}+\frac{q\partial_q}{z(z-q)}-\frac{\partial_z}{z}+\frac{\Delta_1}{z^2}\rt)\bra{\Delta_5}V_4(1)V_3(t)V_2(q)\Phi(z)\ket{\Delta_1}=0.\label{eq:5pt-BPZ}
\end{align}
In 2d Liouville CFT, the correlation function can be expressed in terms of the OPE coefficients $C_{\alpha_i\alpha_j\alpha_k}$ and conformal blocks as 
\begin{align}
\begin{split}
 \bra{\Delta_5}V_4(1)V_3(q)V_2(t)\Phi(z)\ket{\Delta_1}&=\sum_{\theta=\pm}\int{\rm d}\alpha{\rm d}\beta C^{\alpha_{0\theta}}_{\alpha_{2,1}\alpha_0}C^{\alpha}_{\alpha_t\alpha_{0\theta}}C^{\beta}_{\alpha_q\alpha}C_{\alpha_\infty\alpha_1\beta}\\
    &\times\lt|\mathfrak{F}(\Delta_5,\Delta_4,\Delta_d,\Delta_3,\Delta_a,\Delta_2,\Delta_{1\theta},\Delta_{2,1},\Delta_1;z/t,t/q,q)\rt|^2,\\
    \bra{\Delta_5}V_4(1)V_3(q)V_2(t)\ket{\Delta_1}&=\sum_{\theta=\pm}\int{\rm d}\alpha {\rm d}\beta C^{\alpha}_{\alpha_t\alpha_0}C^{\beta}_{\alpha_q\alpha}C_{\alpha_\infty\alpha_1\beta}\cr
    &\times \lt|\mathfrak{F}(\Delta_5,\Delta_4,\Delta_d,\Delta_3,\Delta_a,\Delta_2,\Delta_1;t/q,q)\rt|^2,
\end{split}
\end{align}
where we parameterized the conformal weights as 
\begin{align}
    \Delta_1=\frac{\cQ^2}{4}-\alpha_0^2,\quad \Delta_2=\frac{\cQ^2}{4}-\alpha_t^2,\quad \Delta_3=\frac{\cQ^2}{4}-\alpha_q^2,\quad \Delta_4=\frac{\cQ^2}{4}-\alpha_1^2,\\
    \Delta_5=\frac{\cQ^2}{4}-\alpha_\infty^2,\quad \Delta_a=\frac{\cQ^2}{4}-\alpha^2,\quad \Delta_d=\frac{\cQ^2}{4}-\beta^2,\quad \alpha_{2,1}=-\frac{2b+b^{-1}}{2},\\
     \Delta_{2,1}=\frac{\cQ^2}{4}-\alpha_{2,1}^2=-\frac{1}{2}-\frac{3b^2}{4},\quad \Delta_{1\theta}=\frac{\cQ^2}{4}-\alpha_{0\theta}^2,\quad \alpha_{0\theta}=\alpha_0-\frac{\theta b}{2}.
\end{align}

In the semiclassical limit, $b\to 0$, the conformal blocks become divergent, and leading orders are given by 
\begin{equation}
    \mathfrak{F}(\Delta_5,\Delta_4,\Delta_d,\Delta_3,\Delta_a,\Delta_2,\Delta_1;t/q,q)=t^{\Delta_a-\Delta_2-\Delta_1}q^{\Delta_d-\Delta_3-\Delta_a}e^{b^{-2}F(t/q,q)+\cO(1)},\label{confb-5pt}
\end{equation}
and 
\begin{align}
    &\mathfrak{F}(\Delta_5,\Delta_4,\Delta_d,\Delta_3,\Delta_a,\Delta_2,\Delta_{1\theta},\Delta_{2,1},\Delta_1;z/t,t/q,q)\cr
    &=t^{\Delta_a-\Delta_2-\Delta_{1,\theta}}q^{\Delta_d-\Delta_3-\Delta_a}z^{\frac{b\cQ}{2}+\theta b\a_0}e^{b^{-2}F(t/q,q)+W(z/t,t/q,q)+\cO(b^2)}.\label{leading-confb}
\end{align}
The BPZ equation \eqref{eq:5pt-BPZ} in the semiclassical limit reduces to 
\begin{align}
    \lt(\partial_z^2+\frac{\frac{1}{4}-a_1^2}{(z-1)^2}+\frac{a_0^2+a_q^2+a_t^2+a_1^2-a_\infty^2-3/4-\kappa_5-\mu_5}{z(z-1)}+\frac{\frac{1}{4}-a_t^2}{(z-t)^2}+\frac{\frac{1}{4}-a_q^2}{(z-q)^2}\rt.\cr
    \lt.+\frac{\kappa_5}{z(z-t)}+\frac{\mu_5}{z(z-q)}+\frac{\frac{1}{4}-a_0^2}{z^2}\rt){\cal F}=0,\label{eq:5pt-Schro}
\end{align}
where $a_i:=\lim_{b\to 0}b\alpha_i$ denotes the finite parameters surviving in the semiclassical limit, the solutions to the above equation are normalized as 
\begin{align}
    {\cal F}_\pm(z;t,q):=\lim_{b\rightarrow 0}\frac{\mathfrak{F}(\Delta_5,\Delta_4,\Delta_d,\Delta_3,\Delta_a,\Delta_2,\Delta_{1\pm},\Delta_{2,1},\Delta_1;z/t,t/q,q)}{\mathfrak{F}(\Delta_5,\Delta_4,\Delta_d,\Delta_3,\Delta_a,\Delta_2,\Delta_1;t/q,q)}\cr
    =t^{\mp a_1}z^{\frac{1}{2}\pm a_1}e^{\mp\frac{1}{2}\partial_{a_1}F(t)}\lt(1+\cO(z/t,t/q,q)\rt),\label{semi-reg}
\end{align}
and we have the identifications,
\begin{align}
\begin{split}
    \kappa_5&=\lim_{b\rightarrow 0}b^2t\partial_t\log\mathfrak{F}(\Delta_5,\Delta_4,\Delta_d,\Delta_3,\Delta_a,\Delta_2,\Delta_1;t/q,q),\\
    \mu_5&=\lim_{b\rightarrow 0}b^2q\partial_q\log\mathfrak{F}(\Delta_5,\Delta_4,\Delta_d,\Delta_3,\Delta_a,\Delta_2,\Delta_1;t/q,q).
    \end{split}\label{kappa-mu5}
\end{align}

The connection formula of the special functions can be speculated from the crossing symmetry of the correlation function in the Liouville CFT,
\begin{align}
    &\sum_{\theta=\pm}C^{\alpha_{0\theta}}_{\alpha_{2,1}\alpha_0}C^{\alpha}_{\alpha_t\alpha_{0\theta}}C^{\beta}_{\alpha_q\alpha}C_{\alpha_\infty\alpha_1\beta}\lt|\mathfrak{F}(\Delta_5,\Delta_4,\Delta_d,\Delta_3,\Delta_a,\Delta_2,\Delta_{1\theta},\Delta_{2,1},\Delta_1;z/t,t/q,q)\rt|^2\cr
    &=\sum_{\theta'=\pm} C^{\alpha_{t\theta'}}_{\alpha_{2,1}\alpha_t}C^{\alpha}_{\alpha_0\alpha_{t\theta'}}C^{\beta}_{\alpha_q\alpha}C_{\alpha_\infty\alpha_1\beta}\lt|(1-t)^{\Delta_5-\sum_{i=1}^4\Delta_i-\Delta_{2,1}}\rt|^2\cr
    &\times \lt|\mathfrak{F}\lt(\Delta_5,\Delta_4,\Delta_d,\Delta_3,\Delta_a,\Delta_1,\Delta_{2\theta'},\Delta_{2,1},\Delta_2;\frac{t}{t-1},\frac{t}{t-q},\frac{t-z}{t}\rt)\rt|^2,
    \label{eq:crossing}
\end{align}
where a M\"obius transformation $z\to\frac{z-t}{1-t}$ is also performed on the conformal block on the r.h.s. Graphically, one can write down the connection formula connecting the expansion series around $z\sim 0$ and that around $z\sim t$ for the conformal blocks $\mathfrak{F}$ as 
\begin{align}
    \begin{tikzpicture}
        \draw[ultra thick] (-1,1)--(0,0);
        \draw[ultra thick,dashed] (-1,-1)--(0,0);
        \draw[ultra thick] (0,-1)--(1,0);
        \draw[ultra thick] (2,0)--(0,0);
        \draw[ultra thick] (2,0)--(3,1);
        \draw[ultra thick] (2,0)--(3,-1);
        \draw[ultra thick] (2.5,-0.5)--(3.5,0.5);
        \node at (-1,1) [left] {$\Delta_1^{(0)}$};
        \node at (-1,-1) [left] {$\Delta_{2,1}^{(z)}$};
        \node at (0,-1) [below] {$\Delta_{2}^{(t)}$};
        \node at (3,1) [right] {$\Delta_3^{(q)}$};
         \node at (3.5,0.5) [right] {$\Delta_4^{(1)}$};
        \node at (3,-1) [right] {$\Delta_5^{(\infty)}$};
        \node at (1.5,0) [above] {$\Delta_a$};
        \node at (2.5,-0.5) [left] {$\Delta_d$};
        \node at (0.5,0) [above] {$\Delta_{0,\theta}$};
        \node at (5,0) {$=\tilde{g}(t)\sum_{\theta'}\cA_{\theta\theta'}$};
        \draw[ultra thick] (7,1)--(8,0);
        \draw[ultra thick,dashed] (7,-1)--(8,0);
        \draw[ultra thick] (8,-1)--(9,0);
        \draw[ultra thick] (10,0)--(8,0);
        \draw[ultra thick] (10,0)--(11,1);
        \draw[ultra thick] (10,0)--(11,-1);
        \draw[ultra thick] (10.5,-0.5)--(11.5,0.5);
        \node at (7,1) [left] {$\Delta_2^{(0)}$};
        \node at (7,-1) [left] {$\Delta_{2,1}^{(\frac{z-t}{1-t})}$};
        \node at (8,-1) [below] {$\Delta_{1}^{(\frac{t}{t-1})}$};
        \node at (11,1) [right] {$\Delta_3^{(q')}$};
        \node at (11,-1) [right] {$\Delta_5^{(\infty)}$};
        \node at (9.5,0) [above] {$\Delta_a$};
        \node at (8.5,0) [above] {$\Delta_{2\theta'}$};
        \node at (11.5,0.5) [right] {$\Delta_4^{(1)}$};
    \node at (10.5,-0.5) [left] {$\Delta_d$};
    \end{tikzpicture}
    \label{5-pt-conn}
\end{align}
where $q'=\frac{q-t}{1-t}$ and $\tilde{g}(t)=(1-t)^{\Delta_5-\sum_{i=1}^4\Delta_i-\Delta_{2,1}}$. 
The coefficients $\cA_{\theta\theta'}$ satisfy the following constraint equations,
\begin{align}
    \begin{cases}
        \sum_{\theta}C^{\alpha_{0\theta}}_{\alpha_{2,1}\alpha_0}C^{\alpha}_{\alpha_t\alpha_{0\theta}}C^{\beta}_{\alpha_q\alpha}C_{\alpha_\infty\alpha_1\beta}\cA_{\theta\theta'}\cA^\ast_{\theta\theta'}=C^{\alpha}_{\alpha_{0}\alpha_{t\theta'}}C^{\alpha_{t\theta'}}_{\alpha_{2,1}\alpha_t}C^{\beta}_{\alpha_q\alpha}C_{\alpha_\infty\alpha_1\beta}\\
    \sum_{\theta,\theta'}C^{\alpha_{0\theta}}_{\alpha_{2,1}\alpha_0}C^{\alpha}_{\alpha_t\alpha_{0\theta}}C^{\beta}_{\alpha_q\alpha}C_{\alpha_\infty\alpha_1\beta}\cA_{\theta\theta'}\cA^\ast_{\theta(-\theta')}=0.
    \end{cases}\label{eq:constraint}
\end{align}
They satisfy almost the same constraint equations as $\cM_{\theta\theta'}$ used in the connection formula of the hypergeometric functions (refer to \cite{Bonelli:2022ten} for more details), 
\begin{equation}
    \cM_{\theta\theta'}(\alpha_{0},\alpha_{1};\alpha_{2}):=\frac{\Gamma(-2\theta^{\prime}\alpha_{1})}{\Gamma(\frac{1}{2}+\theta\alpha_{0}-\theta^{\prime}\alpha_{1}+\alpha_{2})}\frac{\Gamma(1+2\theta\alpha_{0})}{\Gamma(\frac{1}{2}+\theta\alpha_{0}-\theta^{\prime}\alpha_{1}-\alpha_{2})},\label{M-def}
\end{equation}
satisfying 
\begin{align}
\begin{cases}
    \sum_{\theta=\pm}C^{\alpha_{0\theta}}_{\alpha_{2,1}\alpha_0} C_{\alpha_\infty\alpha_1\alpha_{0\theta}} \cM_{\theta\theta'}(b\alpha_0,b\alpha_1;b\alpha_\infty)\cM^\ast_{\theta\theta'}(b\alpha_0,b\alpha_1;b\alpha_\infty)=C^{\alpha_{1\theta'}}_{\alpha_{2,1}\alpha_1}C_{\alpha_\infty\alpha_{1\theta'}\alpha_0},\\
    \sum_{\theta,\theta'=\pm }C^{\alpha_{0\theta}}_{\alpha_{2,1}\alpha_0} C_{\alpha_\infty\alpha_1\alpha_{0\theta}}  \cM_{\theta\theta'}(b\alpha_0,b\alpha_1;b\alpha_\infty)\cM^\ast_{\theta(-\theta')}(b\alpha_0,b\alpha_1;b\alpha_\infty)=0.
\end{cases}
\end{align}
Therefore one can soon find a solution to the constraint equations \eqref{eq:constraint} proportional to $\cM_{\theta\theta'}$, 
\begin{equation}
    \cA_{\theta\theta'}\propto\cM(b\alpha_0,b\alpha_t,b\alpha).
\end{equation}
It is always possible to find a new solution to \eqref{eq:constraint} by multiplying an overall phase factor independent of $\theta$ and $\theta'$ to $\cA_{\theta\theta'}$, and one can speculate the phase by comparing the leading order behaviors of both sides in \eqref{eq:crossing}. By using the leading behavior of the 5-pt conformal block plus one degenerate field inserted, \eqref{leading-confb}, i.e. 
\begin{align}
    &\mathfrak{F}(\Delta_5,\Delta_4,\Delta_d,\Delta_3,\Delta_a,\Delta_2,\Delta_{1\theta},\Delta_{2,1},\Delta_1;z/t,t/q,q)\cr
    &=t^{\Delta_a-\Delta_2-\Delta_{1,\theta}}q^{\Delta_d-\Delta_3-\Delta_a}z^{\frac{b\cQ}{2}+\theta b\a_0}\times \lt(1+\cO(z/t,t/q,q)\rt),
\end{align}
the phase factor is thus fixed by 
\begin{equation}
    \cA_{\theta\theta'}=e^{i\pi(\Delta_a-\Delta_1-\Delta_{2,1}-\Delta_2)}\cM_{\theta\theta'}(b\alpha_0,b\alpha_t,b\alpha).
\end{equation}
We note that up to the factor $\tilde{g}(t)$, this expression of the coefficients in the connection formula is the same as that appears in the Heun function. It is then natural to conjecture that the connection formulae for all the special functions as solutions to the ODEs with arbitrary number $n>3$ of regular singularities share the same connection coefficient $\cA_{\theta\theta'}$. To test this conjecture, we apply the above speculated to the study of quasinormal modes of different blackholes and compare the obtained results with well-established numerical methods. In this article, we investigate the cases of $n=5$ and $n=6$. 

In the computation of quasinormal modes, we need to impose the ingoing boundary condition at the horizon and outgoing boundary condition at the infinity (causal boundary), and in terms of the connection coefficients, one needs to solve 
\begin{equation}
    \frac{\cA_{-+}}{\cA_{--}}=0,\label{conn-form-5pt}
\end{equation}
which in the semiclassical limit is equivalent to solve 
\begin{equation}
    \frac{\cM_{-+}(a_0,a_t,a)}{\cM_{--}(a_0,a_t,a)}=0,\quad {\rm with}\ a:=\lim_{b\to 0}b\alpha,\ d:=\lim_{b\to 0}b\beta.\label{conn-form-M}
\end{equation}
The key step is to find the explicit expression of $a$ in terms of other $a_i$ parameters, as $a$ itself does not appear in the ODE \eqref{eq:5pt-Schro} and thus cannot be written directly in terms of blackhole parameters. One can solve for $a$ and $b$ as series expansions in $t/q$ and $q$ from \eqref{kappa-mu5}, by using the explicit expression of the 5-pt conformal block together with the notation $h_i:=b^2\Delta_i$, 
\begin{align}
    &\mathfrak{F}(\Delta_5,\Delta_4,\Delta_d,\Delta_3,\Delta_a,\Delta_2,\Delta_1;t/q,q)=t^{\frac{h_a-h_2-h_1}{b^2}}q^{\frac{h_d-h_3-h_a}{b^2}}\times \cr
    &\lt(1+b^{-2}\frac{(h_a+h_2-h_1)(h_a+h_3-h_b)}{2h_a}\frac{t}{q}+b^{-2}\frac{(h_b+h_3-h_a)(h_4+h_b-h_5)}{2h_b}q+\dots\rt).\cr
\end{align}
At the leading orders, we obtain in the semiclassical limit that 
\begin{align}
	a=&\sqrt{\frac{1}{4}-\kappa_5-h_2-h_1}-\frac{\mu_5(\kappa_5+2h_2)}{2(h_1+h_2+\kappa_5)\sqrt{1-4h_1-4h_2-4\kappa_5}}\frac{t}{q}\cr
	&+\frac{(2h_2+\kappa_5)(h_1+h_4-h_5+h_3+h_2+\kappa_5+\mu_5)}{2\sqrt{1-4h_1-4h_2-4\kappa_5}(h_1+h_2+\kappa_5)}t+\cdots,\label{Matone-5pt-1}\\
	d=&\sqrt{\frac{1}{4}-\mu_5-\kappa_5-h_3-h_2-h_1}\cr
	&+\frac{(2h_3+\mu_5)(h_1+h_4-h_5+h_3+h_2+\kappa_5+\mu_5)}{2\sqrt{1-4h_1-4h_3-4h_2-4\kappa_5-4\mu_5}(h_1+h_3+h_2+\kappa_5+\mu_5)}q+\dots\cr\label{Matone-5pt-2}
\end{align}
The computation for the case of $n=6$ is completely parallel, where we have the corresponding ODE, 
\begin{equation}
    \lt(\partial_z^2+Q_{SW}(z)\rt){\cal G}=0,
\end{equation}
with the convention 
\begin{align}
    Q_{SW}(z)=\frac{\Delta_1}{z^2}+\frac{\Delta_2}{(z-t)^2}+\frac{\Delta_3}{(z-q)^2}+\frac{\Delta_4}{(z-v)^2}+\frac{\Delta_5}{(z-1)^2}\cr
    +\frac{\kappa_6}{z(z-1)}+\frac{\mu_6}{z(z-t)}+\frac{\nu_6}{z(z-q)}+\frac{\theta_6}{z(z-v)},\label{QSW-6pt}\\
    \kappa_6=-\lt(\sum_{i=1}^5\Delta_i+\Delta_{2,1}+z\partial_z-\Delta_6+t\partial_t+q\partial_q+v\partial_v\rt),
\end{align}
and the parameters $\mu_6$, $\nu_6$, $\theta_6$ are related to the 6-pt conformal block as 
\begin{align}
\begin{split}
    \mu_6&=\lim_{b\rightarrow 0}b^2t\partial_t\log\mathfrak{F}(\Delta_6,\Delta_5,\Delta_e,\Delta_4,\Delta_d,\Delta_3,\Delta_a,\Delta_2,\Delta_1;t/q,q/v,v),\\
    \nu_6&=\lim_{b\rightarrow 0}b^2q\partial_q\log\mathfrak{F}(\Delta_6,\Delta_5,\Delta_e,\Delta_4,\Delta_d,\Delta_3,\Delta_a,\Delta_2,\Delta_1;t/q,q/v,v),\\
    \theta_6&=\lim_{b\rightarrow 0}b^2v\partial_v\log\mathfrak{F}(\Delta_6,\Delta_5,\Delta_e,\Delta_4,\Delta_d,\Delta_3,\Delta_a,\Delta_2,\Delta_1;t/q,q/v,v).
    \end{split}\label{kappa-mu6}
\end{align}
The finite parameters $a:=\lim_{b\to0}\sqrt{\frac{1}{4}-b^2\Delta_a}$, $d:=\lim_{b\to0}\sqrt{\frac{1}{4}-b^2\Delta_d}$, and $e:=\lim_{b\to0}\sqrt{\frac{1}{4}-b^2\Delta_e}$ surviving in the semiclassical limit can be found as 
\begin{align}
    &a=\frac{1}{2}\sqrt{1-4\Delta_0-4\Delta_v-4\theta_6}-\frac{(2\Delta_v+\theta_6)\mu_6}{2\sqrt{1-4\Delta_0-4\Delta_v-4\theta_6}(\Delta_0+\Delta_v+\theta_6)}\frac{v}{t}+\dots\\
    &d=\frac{1}{2}\sqrt{1-4\Delta_0-4\Delta_t-4\Delta_v-4\theta_6-4\mu_6}\cr
    &-\frac{2(\Delta_t+\mu_6)\nu_6}{2\sqrt{1-4\Delta_0-4\Delta_t-4\Delta_v-4\theta_6-4\mu_6}(\Delta_0+\Delta_t+\Delta_v+\theta_6+\mu_6)}\frac{t}{q}+\dots\\
    &e=\frac{1}{2}\sqrt{1-4\Delta_0-4\Delta_q-4\Delta_t-4\Delta_v-4\theta_6-4\mu_6-4\nu_6}\cr
    &+\frac{(2\Delta_q+\nu_6)(\Delta_0+\Delta_1-\Delta_\infty+\Delta_q+\Delta_t+\Delta_v+\theta_6+\mu_6+\nu_6)}{2\sqrt{1-4\Delta_0-4\Delta_q-4\Delta_t-4\Delta_v-4\theta_6-4\mu_6-4\nu_6}(\Delta_0+\Delta_q+\Delta_t+\Delta_v+\theta_6+\mu_6+\nu_6)}q\cr
    &+\dots
\end{align}
More generally, one can use the AGT relation proposed in \cite{Alday:2009aq} to compute the higher-order corrections in the 5-pt and 6-pt conformal blocks systematically with the analytic expressions of instanton partition functions of 4d $\cN=2$ quiver SCFTs (more details will be summarized in Appendix \ref{a:quiver}). $a$, $d$, and other conformal weight parameters associated to internal lines can then be found by solving \eqref{kappa-mu5} or \eqref{kappa-mu6} as instanton expansions from the instanton partition functions of quiver gauge theories. 

The connection formulae speculated so far are only for the connection between series expansion solutions around $z\sim 0$ and $z\sim t$. It is also straightforward to mimic the argument in \cite{Bonelli:2022ten} to obtain the connection formulae between other combinations of singular points. Some of them will be presented in Appendix \ref{a:conn}, and will also be tested in this article. 

\section{QNMs in RN-dS blackhole}\label{s:5pt-RNdS}

In this section, we test the connection formulae for ODE with five regular singularities in a simple case, the RN-dS blackhole. In particular, we compare the QNMs obtained from the connection formulae and the results computed from the {\it QNMspectral} package \cite{Jansen:2017oag}. 

The metric of the RN-dS blackhole is given by 
\begin{equation}
    {\rm d}s^2=-f(r){\rm d}t^2+\frac{{\rm d}r^2}{f(r)}+r^2({\rm d}\theta^2+\sin^2\theta{\rm d}\varphi^2),
\end{equation}
with the blackening factor, 
\begin{equation}
    f(r)=1-\frac{2M}{r}+\frac{Q^2}{r^2}-\frac{r^2}{L^2},\quad \Lambda=\frac{3}{L^2},\label{RN5pt:metric1}
\end{equation}
where $M$, $Q$, and $L$ are respectively the blackhole mass, charge and the dS radius of the spacetime. 
We denote four roots of $f(r)=0$ in the ordering $r_-<r_0<r_+<r_c$, where $r_-$ is negative and $r_0$ goes to zero when $Q\rightarrow 0$. Then one can express the blackening factor as 
\begin{equation}
    f(r)=-\frac{1}{r^2L^2}(r-r_-)(r-r_0)(r-r_+)(r-r_c).\label{RN5pt:metric2}
\end{equation}
To solve for the QNMs, one considers the following ODE \cite{Cardoso:2017soq,Aminov:2020yma}, 
\begin{equation}
    \frac{{\rm d}^2\Phi}{{\rm d}r^2_\ast}+\lt(\omega^2-V_{\rm dS}(r)\rt)\Phi=0,\label{KG-RNdS}
\end{equation}
with ${\rm d}r_\ast=\frac{{\rm d}r}{f(r)}$, the potential for perturbations with spin $s$ and angular momentum $l$ given by 
\begin{equation}
    V_{\rm dS}(r)=f(r)\lt(\frac{l(l+1)}{r^2}+(1-s^2)\lt(\frac{f'(r)}{r}+\frac{s^2\Lambda}{6}\rt)\rt),
\end{equation}
and the boundary conditions, 
\begin{equation}\label{eq:bdy-dS}
    \phi(r)\sim\begin{cases}
e^{-i\omega r_{\ast}}\propto (r-r_+)^{-\frac{i\omega r_+^2L^2}{(r_+-r_-)(r_+-r_0)(r_c-r_+)}} & r\to r_{+}\\
e^{i\omega r_{\ast}}\propto (r-r_c)^{-\frac{i\omega r_c^2L^2}{(r_c-r_-)(r_c-r_0)(r_c-r_+)}}& r\to r_c
\end{cases}.
\end{equation}
The ODE for scalar perturbations $s=0$ has five regular singularities for $s=0$ at $r=r_-,r_0,r_+,r_c,\infty$, and to utilize the connection formulae guessed in the previous section, we need to map the ODE \eqref{KG-RNdS} to the 5-pt BPZ equation in the semiclassical limit, \eqref{eq:5pt-Schro}. There are many ways to map the above five singularities to $z=0,t,q,1,\infty$, and we take one of them to analyze the converging behavior of the numerical results. 

Let us perform the coordinate transformation, 
\begin{equation}
    z=\frac{r_--r_0}{r_--r_+}\frac{r-r_+}{r-r_0},\label{coorTran}
\end{equation}
which maps $r=r_0$ to the infinity, $r=r_-$ to $z=1$, $r=r_+$ to $z=0$, $r=r_c$ to $z=t$ with
\begin{equation}
    t=\frac{r_--r_0}{r_--r_+}\frac{r_c-r_+}{r_c-r_0},
\end{equation}
and the infinity to $z=q$,
\begin{equation}
    q=\frac{r_0-r_-}{r_+-r_-}.
\end{equation}
Bringing the ODE \eqref{KG-RNdS} after the coordinate transform \eqref{coorTran} to the Schr\"odinger form, one can map the potential to the following form, 
\begin{align}
    Q_{\rm SW}(z)=\frac{\Delta_0}{z^{2}}+\frac{\Delta_1}{(z-1)^{2}}+\frac{\Delta_t}{(z-t)^{2}}+\frac{\eta_5}{z(z-1)}+\frac{\kappa_5}{z(z-t)}
    +\frac{\Delta_q}{(z-q)^2}+\frac{\mu_5}{z(z-q)},\label{RN5pt:QSW}
\end{align}
with the following dictionary, which will be referred to as Dictionary 1 in this article: 
\begin{align}
    &\Delta_0=\frac{1}{4}+\frac{9r_+^4\omega^2}{(r_0-r_+)^2(r_c-r_+)^2(r_+-r_-)^2\Lambda^2},\quad \Delta_1=\frac{1}{4}+\frac{9r_-^4\omega^2}{(r_0-r_-)^2(r_c-r_-)^2(r_+-r_-)^2\Lambda^2},\\
   &\Delta_t=\frac{1}{4}+\frac{9r_c^4\omega^2}{(r_c-r_0)^2(r_c-r_-)^2(r_c-r_+)^2\Lambda^2},\quad \Delta_q=-2,\quad\mu_5=-\frac{-3r_0+r_c+r_-+r_+}{r_0-r_+}\\
    &\kappa_5=-\frac{18r_c^3(-2r_-r_++r_c(r_-+r_+))}{(r_0-r_c)(r_c-r_-)^3(r_0-r_+)(r_c-r_+)^2\Lambda^2}\omega^2-\frac{6l+6l^2+(r_0-r_c)(2r_c-r_--r_+)\Lambda}{2(r_c-r_-)(r_0-r_+)\Lambda}\\
    &\eta_5=-\frac{18r_-^3(r_c(r_--2r_+)+r_-r_+)}{(r_0-r_-)(-r_c+r_-)^3(r_0-r_+)(r_--r_+)^2\Lambda^2}\omega^2+\frac{6l+6l^2-(r_0-r_-)(r_c-2r_-+r_+)\Lambda}{2(r_c-r_-)(r_0-r_+)\Lambda}.\label{RN5pt:dic}
\end{align}
In Table \ref{RN5pt:QNMs-1} and Table \ref{RN5pt:QNMs-2}, we list the quasinormal modes obtained from the connection formula \eqref{conn-form-5pt} compared with the results from the {\it QNMspectral} package for various combinations of parameters. It is known that there are in total three families of QNMs \cite{Cardoso:2017soq}: photon sphere modes (PS modes), dS modes, and near-extremal modes. In this parameter region, we focus on, one observes two families: purely imaginary QNMs corresponding to dS modes and complex ones as PS modes. 

One observes in the computation in Dictionary 1 that when $\Lambda M^2$ is large and close to its critical value $\Lambda_c M^2=\frac{1}{9}$ for $Q=0$, the numerical results from the connection formula become better. We also remark that compared to PS modes, it is more difficult to find the purely imaginary dS modes. In the calculation of the connection formula, we approximate $a$ by its instanton expansion up to some order. For the PS modes with $\Lambda M^2\geq 0.1$, we expanded $a$ as 
\begin{align}
    a\approx a_{00}+a_{10}\frac{t}{q}+a_{11}t+a_{20}\frac{t^2}{q^2}+a_{21}\frac{t^2}{q}+a_{30}\frac{t^3}{q^3}+a_{22}t^2+a_{31}\frac{t^3}{q^2}+a_{13}tq^2,\label{expansion-a}
\end{align}
where the coefficients $a_{ij}$ are determined from \eqref{kappa-mu5} as explained in \eqref{Matone-5pt-1} and we note that as e.g. $a_{01}=0$, some of the expansion coefficients do not appear above. We remark that many coefficients $a_{ij}$ with $i<j$ are vanishing. This is due to the fact that $a$ is computed from the first relation in \eqref{kappa-mu5}, and because of the asymmetry between $t/q$ and $q$ introduced by the $t$-derivative $t\partial_t$, one can directly see that $a_{0i}$ vanishes for all $i$ due to the $t$-derivative. Interestingly $a_{12}$ further vanishes in a very non-trivial way, but we are not sure whether there is any profound physical reason behind it. To work out the dS modes and PS modes for $\Lambda M^2=0.08, 0.06$, we further needed to include a higher-order term $a_{40}t^4/q^4$, and at $\Lambda M^2=0.08$ we had to add terms up to $a_{50}t^5/q^5$ to reproduce the numerical result of the dS mode $\omega\simeq -0.16i$.

\begin{table}
    \centering
    \begin{tabular}{|c|c|c|c|}\hline
      & connection formula & {\it QNMspectral} package &  \\\hline
      $\Lambda M^2=0.111$ & $\pm0.0167996-0.00634621i$ & $\pm0.0167995-0.0063458i$ & $t=0.049652$\\
      $s=0$, $l=1$ & $\pm0.0167758-0.0190387i$ & $\pm0.0167764-0.0190376i$ & $q=0.675981$\\
      $Q/M=0.1$ & $\pm0.0167281-0.0317313i$ & $\pm0.0167299-0.0317302i$ & $ $\\
       & $\pm0.016656-0.0444243i$ & $\pm0.01666-0.04442i$ & $ $\\
      \hline
      $\Lambda M^2=0.105$ & $\pm0.0621883-0.0237516i$ & $\pm0.0619707-0.0235847i$  & $t=0.173325$\\
      $s=0$, $l=1$ & $\pm0.0608906-0.0714109i$ & $\pm0.0609454-0.0709165i$ & $q=0.700287$\\
      $Q/M=0.1 $ & $\pm0.0580448-0.119777i$ & $\pm0.0587513-0.1189427i$ & $ $\\
      \hline
      $\Lambda M^2=0.1$ & $\pm0.0835771-0.0322306i$ & $\pm0.0829927-0.0317380i$ & $t=0.227257$\\
      $s=0$, $l=1$ & $\pm0.0805761-0.0972907i$ & $\pm0.0808842-0.0958549i$ & $q=0.712344$\\
      $Q/M=0.1 $ & $\pm0.074882-0.166325i$ & $\pm0.077409-0.162856i$ & $ $\\
       & $-0.17286i$ & $-0.18179i$ & $ $\\
      \hline
      $\Lambda M^2=0.08$ & $ \pm0.14665-0.0599754i$ & $\pm0.1413191-0.0542971i $ & $t=0.369929$\\
      $s=0$, $l=1$ & $\pm0.157417-0.168172i $ & $\pm0.1379974-0.1658399i $ & $q=0.749023$\\
      $Q/M=0.1$ & $-0.175212i $ & $-0.1626846i $ &  \\
      \hline
      $\Lambda M^2=0.06$ & $\pm0.214369-0.0735466i$ & $\pm0.186112-0.0702313i$ & $t=0.478886$\\
      $s=0$, $l=1$ & $\pm0.204225-0.216968i$ & $\pm0.181882-0.212352i$ & $q=0.782069$\\
      $Q/M=0.1$ &  $-0.151311i$ & $-0.141007i$ &  \\
      \hline
      \end{tabular}
    \caption{Comparison of QNMs in the RN-dS blackhole at $Q/M=0.1$ and $l=1$ obtained from the 5pt connection formula with numerical data obtained from the {\it QNMspectral} package.}
    \label{RN5pt:QNMs-1}
\end{table}

\begin{table}
    \centering
    \begin{tabular}{|c|c|c|c|}\hline
      & connection formula & {\it QNMspectral} package &  \\\hline
      $\Lambda M^2=0.111$ & $\pm0.0305746-0.00634212i$ & $\pm0.0304379-0.0063406i$ & $t=0.049652$\\
      $s=0$, $l=2$ & $\pm0.030565-0.0190264i$ & $\pm0.0304295-0.0190219i$ & $q=0.675981$\\
      $Q/M=0.1$ & $\pm0.0305458-0.0317108i$ & $\pm0.0304126-0.0317033i$ & $ $\\
       & $\pm0.0305168-0.0443954i$ & $\pm0.0304-0.0444i$ & $ $\\
      \hline
      $\Lambda M^2=0.105$ & $\pm0.113702-0.0234062i$ & $\pm0.1117754-0.0233364i$  & $t=0.173325$\\
      $s=0$, $l=2$ & $\pm0.113268-0.0702417i$ & $\pm0.1113831-0.0700268i$ & $q=0.700287$\\
      $Q/M=0.1 $ & $\pm0.112382-0.117152i$ & $\pm0.11059-0.11677i$ & $ $\\
       & $\pm0.111011-0.164211i $ & $\pm0.11-0.16i $ & $ $\\
      \hline
      $\Lambda M^2=0.1$ & $\pm0.152581-0.0313331i$ & $\pm0.1490722-0.0311683i$ & $t=0.227257$\\
      $s=0$, $l=2$ & $\pm0.151614-0.0940824i$ & $\pm0.1481920-0.0935729i$ & $q=0.712344$\\
      $Q/M=0.1 $ & $\pm0.149651-0.157119i$ & $\pm0.146406-0.156207i$ & $ $\\
      & $\pm0.146661-0.220793i$ & $\pm0.14-0.22i$ & $ $\\
      \hline
      $\Lambda M^2=0.08$ & $\pm0.25988-0.0531617i $ & $\pm0.2489838-0.0521598i $ & $t=0.369929$\\
      $s=0$, $l=2$ & $\pm0.255992-0.159984i $ & $\pm0.2457421-0.1570283i $ & $q=0.749023$\\
      $Q/M=0.1$ &$\pm0.249071-0.268297i$ &$\pm0.2398-0.2636i$ & \\
      \hline
      $\Lambda M^2=0.06$ & $\pm0.344403-0.0712959i$ & $\pm0.3212215-0.0670125i$ & $t=0.478886$\\
      $s=0$, $l=2$ & $\pm0.334089-0.214295i$ & $\pm0.3154032-0.2021392i$ & $q=0.782069$\\
       $Q/M=0.1$ & $-0.284827i$ & $-0.28i$ &  \\
      \hline
      \end{tabular}
    \caption{Comparison of QNMs in the RN-dS blackhole at $Q/M=0.1$ and $l=2$ obtained from the 5pt connection formula with numerical data obtained from the {\it QNMspectral} package.}
    \label{RN5pt:QNMs-2}
\end{table}

\begin{table}
    \centering
    \begin{tabular}{|c|c|c|c|}\hline
      & connection formula & {\it QNMspectral} package &  \\\hline
      $\Lambda M^2=0.1$ & $\pm0.344-0.0609i$ & $\pm0.4197010-0.0613848i$ & $t=-0.0644669$\\
      $s=0$, $l=2$ & $\pm0.407 - 0.185i$ & $\pm0.4140070-0.1846877i$ & $q=0.270243$\\
      $Q/M=0.999 $ & $\pm0.403 - 0.394i$ & $\pm0.4027-0.3096i$ & $ $\\
       & $-0.392i$ & $-0.365i$ & \\
      & $-0.587i$ & $-0.565i$ & \\
      \hline
      $\Lambda M^2=0.06$ & $\pm0.491- 0.0829i$ & $\pm0.5106515-0.0740624i$ & $t=-0.0374809$\\
      $s=0$, $l=2$ & $\pm0.469 - 0.204i$ & $\pm0.5008522-0.2232539i$ & $q=0.336537$\\
      $Q/M=0.999$ & $-0.334i$ & $-0.2828i$& \\
       & $-0.428i$ & $-0.4456i$ &  \\
      \hline
      $\Lambda M^2=0.03$ & $\pm0.381-0.077i$ & $\pm0.570717-0.082008i$ & $t=-0.0194778$\\
      $s=0$, $l=2$ & $-0.191i$ & $-0.20i$ & $q=0.390932$\\
      $Q/M=0.999$ & $-0.341i$ & $-0.33i$ &  \\
      \hline
      $\Lambda M^2=0.06$ & $\pm0.311 - 0.0702i$ & $\pm0.302392-0.076006i$ & $t=-0.0374809$\\
      $s=0$, $l=1$ & $-0.110i$ & $-0.141i$ & $q=0.336537$\\
      $Q/M=0.999$ & $-0.332i$ & $-0.297i$& \\
      \hline
      \end{tabular}
    \caption{Low-order computations of QNMs of RN-dS blackhole at $Q/M=0.999$ obtained from the 5pt connection formula vs numerical data obtained from the {\it QNMspectral} package.}
    \label{RN5pt:QNMs-3}
\end{table}

However, Dictionary 1 does not work for cases with relatively large $Q/M$, as one can easily check that $t$ becomes large and the instanton expansion \eqref{expansion-a} breaks down for $a$. In such a situation, one can apply another useful coordinate transformation, 
\begin{equation}
    z=\frac{r-r_+}{r-r_-}.
\end{equation}
It maps $r=r_-$ to the infinity, $r=r_+$ to $z=0$, $r=r_0$ to $z=t$ with 
\begin{equation}
    t=\frac{r_0-r_+}{r_0-r_-}<0,
\end{equation}
and $r=r_c$ to $z=q$ with
\begin{equation}
    q=\frac{r_c-r_+}{r_c-r_-}>0.
\end{equation}
Then we obtain the following dictionary, which will be referred to as Dictionary 2, for the physical parameters in the semiclassical BPZ equation \eqref{eq:5pt-Schro}, 
\begin{align}
    &\Delta_0=\frac{1}{4}+\frac{9r_+^4\omega^2}{(r_0-r_+)^2(r_c-r_+)^2(r_+-r_-)^2\Lambda^2},\quad \Delta_t=\frac{1}{4}+\frac{9r_0^4\omega^2}{(r_0-r_c)^2(r_0-r_-)^2(r_0-r_+)^2\Lambda^2},\\
   &\Delta_q=\frac{1}{4}+\frac{9r_c^4\omega^2}{(r_0-r_c)^2(r_c-r_-)^2(r_c-r_+)^2\Lambda^2},\quad \Delta_1=-2,\quad\eta_5=\frac{r_0+r_c-3r_-+r_+}{r_+-r_-}\\
    &\kappa_5=\frac{18r_0^3(-2r_cr_++r_0(r_c+r_+))}{(r_0-r_c)^3(r_0-r_-)(r_0-r_+)^2(r_--r_+)^2\Lambda^2}\omega^2+\frac{-6l-6l^2+(r_0-r_m)(2r_0-r_c-r_+)\Lambda}{2(r_0-r_c)(r_--r_+)\Lambda}\\
    &\mu_5=\frac{-18r_0r_c^4+36r_0r_c^3r_+-18r_c^4r_+}{(r_0-r_c)^3(r_c-r_-)(r_c-r_+)^2(r_--r_+)\Lambda^2}\omega^2+\frac{6l+6l^2+(r_c-r_-)(r_0-2r_c+r_+)\Lambda}{2(r_0-r_c)(r_--r_+)\Lambda}.\label{RN5pt:dic-2}
\end{align}
It is then possible to use the connection formula connecting $z\sim 0$ and $z\sim q$, presented in \eqref{eq:conn-0-q}, to compute the QNMs. Some numerical results are listed in Table \eqref{RN5pt:QNMs-3} with low-order approximation of $a\approx a_{00}+a_{10}\frac{t}{q}+a_{11}t$, $d\approx d_{00}+d_{01}q+d_{11}t$ and $\partial_a F$. One can observe that in Dictionary 2, it is easier to access purely imaginary modes than PS modes. By comparing with the results shown in \cite{Cardoso:2017soq} e.g. at $\Lambda M^2=0.06$, $l=1$, we see that the purely imaginary modes found by the {\it QNMspectral} package correspond to the family of dS modes, and in fact, it overlooked some near-extremal modes that become more dominant in this case. These near-extremal modes, fortunately, can be accessed via the connection formula approach. We will try to explain the phenomenon observed above in the next section, and some analytic results of the near-extremal modes will also be presented.

\section{Analytic results}\label{s:ana}

More interestingly, using the method of the connection formula, one can obtain some analytic results about the quasinormal modes as some series expansion. In this section, we present the methodology on how to extract the analytic expressions and more concretely, we give the expression of PS modes and near-extremal modes as series expansions.

In Dictionary 1 of the RN-dS blackhole, as discussed before, when $Q/M$ is small and $\Lambda M^2$ is closed to the critical value $1/9$, $t$ becomes small and the expansion over $\lt(1-9\Lambda M^2\rt)$ and $Q/M$ will be effective. More explicitly, at the leading orders of these expansion parameters, we have 
\begin{align}
    r_c=3+\sqrt{3}(1-9\Lambda)^{\frac{1}{2}}+\frac{4}{3}(1-9\Lambda)+\frac{Q^2}{2\sqrt{3}(1-9\Lambda)^{\frac{1}{2}}}+{\cal O}\lt((1-9\Lambda)^{\frac{3}{2}},Q^2(1-9\Lambda)^{\frac{1}{2}}\rt),\cr
    r_+=3-\sqrt{3}(1-9\Lambda)^{\frac{1}{2}}+\frac{4}{3}(1-9\Lambda)-\frac{Q^2}{2\sqrt{3}(1-9\Lambda)^{\frac{1}{2}}}+{\cal O}\lt((1-9\Lambda)^{\frac{3}{2}},Q^2(1-9\Lambda)^{\frac{1}{2}}\rt),\cr
    r_-=-6-\frac{8}{3}(1-9\Lambda)-\frac{Q^2}{18}+{\cal O}\lt((1-9\Lambda)^{2},Q^2(1-9\Lambda)^{\frac{1}{2}}\rt),\quad 
    r_0=-\frac{Q^2}{6}+{\cal O}\lt(Q^4\rt),
\end{align}
and 
\begin{align}
    t=\frac{4\sqrt{3}}{9}(1-9\Lambda)^{\frac{1}{2}}+\frac{2\sqrt{3}}{27}\frac{Q}{(1-9\Lambda)^{\frac{1}{2}}}+\dots,\\
    q=-\frac{2}{3}+\frac{2\sqrt{3}}{27}(1-9\Lambda)^{\frac{1}{2}}+\dots,
\end{align}
Here and in the following, we set $M\equiv 1$ without losing generality for the simplicity of notations. 
Following the above expansion, we obtain at the leading orders, 
\begin{align}
    a_0=\frac{3ir_+^2\omega}{(r_+-r_0)(r_c-r_+)(r_+-r_-)\Lambda}=\frac{9\sqrt{3}i\omega}{6(1-9\Lambda)^{\frac{1}{2}}+\frac{Q^2}{(1-9\Lambda)^{\frac{1}{2}}}}+\dots,\\
    a_t=\frac{3ir_c^2\omega}{(r_c-r_0)(r_c-r_+)(r_c-r_-)\Lambda}=\frac{9\sqrt{3}i\omega}{6(1-9\Lambda)^{\frac{1}{2}}+\frac{Q^2}{(1-9\Lambda)^{\frac{1}{2}}}}+\dots,
\end{align}
and 
\begin{align}
    a=\frac{1}{2}\lt(1-4\ell-4\ell^2-48\omega^2\rt)^{\frac{1}{2}}+{\cal O}\lt((1-9\Lambda)^{\frac{1}{2}},Q^2(1-9\Lambda)^{-\frac{1}{2}}\rt).
\end{align}
One can check via numerical computations that in Dictionary 1, the PS modes correspond to the poles of the following $\Gamma$-function in the factor $\cM$ in \eqref{conn-form-M},  
\begin{equation}
    \Gamma^{-1}\lt(\frac{1}{2}-a_0-a_t+a\rt)=0,
\end{equation}
or more concretely we have 
\begin{equation}
    \frac{1}{2}-a_0-a_t+a=-n,\quad n\in\mathbb{N}.
\end{equation}
By assuming that both $(1-9\Lambda)^{\frac{1}{2}}$ and $Q^2(1-9\Lambda)^{-\frac{1}{2}}$ are small (of the same order), the PS quasinormal modes at the leading order of these expansion parameters are given by 
\begin{equation}
    \omega_{\rm PS}=\frac{6(1-9\Lambda)^{\frac{1}{2}}+Q^2(1-9\Lambda)^{-\frac{1}{2}}}{18\sqrt{3}}\lt[\pm\frac{1}{2}\sqrt{4\ell^2+4\ell-1}-\lt(n+\frac{1}{2}\rt)i\rt]+\dots,\quad n\in\mathbb{N}.\label{PS-ana}
\end{equation}
One can compare the above analytic expression (see Table \ref{analytic-1}) with the complex QNMs in Table \ref{RN5pt:QNMs-1} and \ref{RN5pt:QNMs-2}, and see that they match pretty well especially around $\Lambda M^2\approx 0.08$ (for $\ell=2$, this extends to $\Lambda M^2\approx 0.06$) to $\Lambda M^2\approx 0.1$ for $Q/M=0.1$. When $\Lambda M^2$ approaches its critical value $1/9$, $Q^2(1-9\Lambda)^{-\frac{1}{2}}$ will become larger, so the leading-order approximation also gets worse around $\Lambda M^2\approx 0.111$ for $Q/M=0.1$. We note that even though we only worked out the analytic expression at the leading order, its numerical precision is still good enough to be compared with our brute-force connection formula computation at high order. This is due to the fact that each coefficient in the instanton expansion of the connection formula is a complicated function of $\Lambda$ and $Q$, and the convergence of the series becomes rather chaotic when we increase the number of terms included. 

We further remark that the analytic expansion in terms of $(1-9\Lambda)^{\frac{1}{2}}$ and $Q^2(1-9\Lambda)^{-\frac{1}{2}}$ is a re-expansion of the instanton series in the region $|t/q|\ll |q|<1$. This new series expansion works because the instanton-expansion coefficients before terms involving only $q$, i.e. $a_{0i}$, are all vanishing as discussed before. 

The analytic expression of the purely imaginary dS modes, on the other hand, cannot be obtained in Dictionary 1 from some pole of the $\Gamma$-functions, as it cannot be worked out perturbatively from the expansion in $(1-9\Lambda)^{\frac{1}{2}}$ and $Q^2(1-9\Lambda)^{-\frac{1}{2}}$. One would expect to apply similar tricks for dS modes in the parameter region of small $\Lambda M^2$, and thus we need to consider maps that give $t\propto \Lambda$.

\begin{table}
    \centering
    \begin{tabular}{|c|c|c|c|}\hline
      & connection formula & {\it QNMspectral} package & leading-order analytic  \\\hline
      $\Lambda M^2=0.1$ & $\pm0.0835771-0.0322306i$ & $\pm0.0829927-0.0317380i$ & $\pm0.0818494 - 0.0309362i$\\
      $s=0$, $l=1$ & $\pm0.0805761-0.0972907i$ & $\pm0.0808842-0.0958549i$ & $\pm0.0818494 - 0.0928085i$\\
      & $\pm0.074882-0.166325i$ & $\pm0.077409-0.162856i$ & $\pm 0.0818494 - 0.154681i$\\
      \hline
      $\Lambda M^2=0.08$ & $ \pm0.14665-0.0599754i$ & $\pm0.1413191-0.0542971i $ & $\pm0.135517 - 0.0512206i$\\
      $s=0$, $l=1$ & $\pm0.157417-0.168172i$ & $\pm0.1379974-0.1658399i $ & $\pm0.135517 - 0.153662i$\\
      \hline
      $\Lambda M^2=0.1$ & $\pm0.152581-0.0313331i$ & $\pm0.1490722-0.0311683i$ & $\pm0.148365 - 0.0309362i$\\
      $s=0$, $l=2$ & $\pm0.151614-0.0940824i$ & $\pm0.1481920-0.0935729i$ & $\pm0.148365 - 0.0928085i$\\
         & $\pm0.149651-0.157119i$ & $\pm0.146406-0.156207i$ & $\pm0.148365 - 0.154681i $\\
      & $\pm0.146661-0.220793i$ & $\pm0.14-0.22i$ & $\pm0.148365 - 0.216553i $\\
      \hline
      $\Lambda M^2=0.08$ & $\pm0.25988-0.0531617i $ & $\pm0.2489838-0.0521598i $ & $\pm0.245645 - 0.0512206i$\\
      $s=0$, $l=2$ & $\pm0.255992-0.159984i $ & $\pm0.2457421-0.1570283i $ & $\pm0.245645 - 0.153662i$\\
      &$\pm0.249071-0.268297i$ &$\pm0.2398-0.2636i$ & $\pm0.245645 - 0.256103i$\\
      \hline
      \end{tabular}
    \caption{Comparison of QNMs in the RN-dS blackhole at $Q/M=0.1$ obtained from the 5pt connection formula with numerical data obtained from the {\it QNMspectral} package.}
    \label{analytic-1}
\end{table}

In Dictionary 2, the instanton expansion parameter $t$ becomes small in the near-extremal region. From the numerical computation, we see that the dS modes seem to come from the poles of the following $\Gamma$-function appearing in the prefactor $\cM$ in the connection formula \eqref{eq:conn-0-q},
\begin{equation}
    \Gamma^{-1}\lt(\frac{1}{2}+a-a_q-d\rt)=0,
\end{equation}
and the near-extremal modes correspond to 
\begin{equation}
    \Gamma^{-1}\lt(\frac{1}{2}+a-a_0-a_t\rt)=0.
\end{equation}
We remark that this is an almost parallel conclusion to the method proposed in \cite{Lei:2023mqx} to identify the accelerating and near-extremal modes in the charged C-metric (accelerating blackhole). The dS modes stay almost the same when varying $Q/M$, but the near-extremal modes suddenly become very small and accessible when $Q/M\simeq 1$. One can expand all physical parameters around the extremal point $Q^2/M^2=1+\frac{\Lambda^2 M^4}{6}-\delta^2$ with small $\delta$ and $\Lambda$. Then we have 
\begin{equation}
    t=-2\sqrt{\frac{\Lambda}{3}}\delta+\dots,\quad q=\frac{1}{2}-(1+\delta/2)\sqrt{\frac{\Lambda}{3}}+\dots
\end{equation}
and 
\begin{align}
    r_+=1+\delta+\frac{1}{12}\frac{\Lambda}{\delta}+{\cal O}(\Lambda,\delta^2),\quad r_0=1-\delta-\frac{1}{12}\frac{\Lambda}{\delta}+{\cal O}(\Lambda,\delta^2),\\
    r_c=\sqrt{\frac{3}{\Lambda}}-1+{\cal O}(\sqrt{\Lambda},\delta^2),\quad r_-=-\sqrt{\frac{3}{\Lambda}}-1+{\cal O}(\sqrt{\Lambda},\delta^2).
\end{align}
To obtain the leading order analytic expression for the dS modes, we use that 
\begin{align}
    a=l+\frac{1}{2}+\dots,\quad d=\frac{1}{2}\sqrt{\frac{3}{\Lambda}}i\omega+\dots,\quad a_q=\frac{1}{2}\sqrt{\frac{3}{\Lambda}}i\omega+\dots,
\end{align}
and this gives 
\begin{equation}
    \omega_{\rm dS}=-i(l+n+1)\sqrt{\frac{\Lambda}{3}}+{\cal O}(\delta,\Lambda),\quad n\in\mathbb{N}.\label{dS-ana}
\end{equation}
Similarly, for the near-extremal modes, further from 
\begin{align}
    a_0=\frac{i}{2\delta}\omega+i\omega+\dots,\quad a_t=\frac{i}{2\delta}\omega-i\omega+\dots,
\end{align}
one obtains the leading-order behavior of the near-extremal modes as 
\begin{equation}
    \omega_{\rm NE}=-i(l+n+1)\delta +{\cal O}(\delta^2,\Lambda \delta).\label{NE-ana}
\end{equation}
The analytic expressions obtained at the leading order match perfectly with the ones conjectured in \cite{Cardoso:2017soq} from numerical studies. We note that the advantage of our approach is that one can easily go beyond to obtain higher-order corrections for these modes if necessary.

\section{Application to non-scalar Perturbations in RN-dS blackhole }\label{s:non-s}

In the previous sections, we showed that the connection formula approach can provide solid access to both the numerical and analytical QNM analysis for scalar perturbations of RN-dS blackholes. In this section, we further demonstrate that a similar method is applicable to the study of non-scalar perturbations in RN-dS blackholes. 

When $s>0$, the QNM equation \eqref{KG-RNdS} generically possesses 6 regular singularities. One can still implement the same coordinate transform as that used in finding Dictionary 1, 
\begin{align}
    z=\frac{(r_--r_0)(r-r_+)}{(r_--r_+)(r-r_0)}
\end{align}
but in the current case, there will be an additional regular singularity at $r=0$, which is mapped to $z=v$, 
\begin{align}
    v=\frac{-r_0r_++r_-r_+}{r_0(r_--r_+)}.
\end{align}
Bringing the equation to the Sch\"odinger form and comparing with the potential \eqref{QSW-6pt}, we obtain the following dictionary. 

\begin{align}
    \Delta_0=\frac{1}{4}+\frac{9\ r_+^4\omega^2}{(r_0-r_+)^2(r_c-r_+)^2(r_--r_+)^2\Lambda^2},\quad\Delta_t=\frac{1}{4}+\frac{9\ r_c^4\omega^2}{(r_0-r_c)^2(r_c-r_-)^2(r_c-r_+)^2\Lambda^2}\cr
     \Delta_q=\frac{1}{4}(-4+5s^2-s^4),\quad\Delta_v=-2s^2,\quad\Delta_1=\frac{1}{4}+\frac{9\ r_-^4\omega^2}{(r_0-r_-)^2(r_c-r_-)^2(r_--r_+)^2\Lambda^2}
\end{align}
\begin{align}
   \kappa_6=-\frac{18r_-^3(r_c(r_--2r_+)+r_-r_+)}{(r_0-r_-)(-r_c+r_-)^3(r_0-r_+)(r_--r_+)^2\Lambda^2}\omega^2+\frac{6l+6l^2-(r_0-r_-)(r_c-2r_-+r_+)\Lambda}{2(r_c-r_-)(r_0-r_+)\Lambda}\cr
   +\frac{-2r_0(r_c-r_-)(r_--r_+)+r_-(2r_c(r_--r_+)+r_-(-3r_-+2r_+))}{2r_-(-r_c+r_-)(r_0-r_+)}s^2-\frac{r_-^2}{2(r_c-r_-)(r_0-r_+)}s^4,
\end{align}
\begin{align}
    \mu_6=-\frac{18r_c^3(-2r_-r_++r_c(r_-+r_+))}{(r_0-r_c)(r_c-r_-)^3(r_0-r_+)(r_c-r_+)^2\Lambda^2}\omega^2-\frac{6l(l+1)+(r_0-r_c)(2r_c-r_--r_+)\Lambda}{2(r_c-r_-)(r_0-r_+)\Lambda}\cr
    +\frac{2r_0(r_c-r_-)(r_c-r_+)+r_c(-3r_c^2-2r_-r_++2r_c(r_-+r_+))}{2r_c(r_c-r_-)(r_0-r_+)}s^2+\frac{r_c^2}{2(r_c-r_-)(r_0-r_+)}s^4,
\end{align}
\begin{align}
    &\nu_6=\frac{(-1+s^2)(r_0(-6+s^2)-(r_c+r_-+r_+)(-2+s^2))}{2(r_0-r_+)},\\
    &\theta_6=\frac{(r_0r_cr_--3r_cr_-r_++r_0(r_c+r_-)r_+)s^2}{r_cr_-(r_0-r_+)}.
\end{align}
Substituting the above dictionary into the connection formula \eqref{conn-form-M} (taking the same form for both 5-pt and 6-pt BPZ equations), we obtain some numerical results for the PS modes, and some of them are listed in Table \ref{RN6pt:QNMs}. We remark that the convergence of the instanton expansion becomes worse in the 6-pt case than before. 

\begin{table}
    \centering
    \begin{tabular}{|c|c|c|c|}\hline
      & Conformal block data & Numerical data &  \\\hline
      $\Lambda M^2=0.111$ & $\pm0.0213467-0.00634314i$ & $\pm0.0167890-0.0063367i$ & $t=0.049652$\\
      $s=1$,$l=1$ & $\pm0.0213316-0.0190294i$ & $\pm0.0167797-0.0190100i$ & $q=0.675981$\\
      $Q/M=0.1$ & $\pm0.0213014-0.0317157i$ & $\pm0.016761-0.031683i$ & $ $\\
      $ $ & $\pm0.0212561-0.044402i$ & $\pm0.017-0.044i$ & $ $\\
      \hline
      $\Lambda M^2=0.111$ & $\pm0.0334224-0.00634314i$ & $\pm0.0304309-0.0063379i$ &\\
      $s=1$,$l=2$ & $\pm0.0334128-0.0190294i$ & $\pm0.0304250-0.0190128$ &\\
      & $\pm0.0333935-0.0317157i$ & $\pm0.0304131-0.0316880i$ & $ $\\
      $ $ & $\pm0.0333646-0.044402i$ & $\pm0.030-0.044i$ & $ $\\
      \hline
      $\Lambda M^2=0.105$ & $\pm0.0815791-0.0235535i$ & $\pm0.0614357-0.0231477i$ & $t=0.173325$\\
      $s=1$,$l=1$ & $\pm0.080743-0.0706604i$ & $\pm0.0609648-0.0694504i$ & $q=0.700287$\\
      & $\pm0.0790442-0.117767i$ & $\pm0.060001-0.115778i$ & $ $\\
      $ $ & $\pm0.0764252-0.164874i$ & $\pm0.06-0.16i$ & $ $\\
      \hline
      $\Lambda M^2=0.105$ & $\pm0.12771-0.0235535i$ & $\pm0.1114268-0.0231930i$ &\\
      $s=1$,$l=2$ & $\pm0.127178-0.0706604i$ & $\pm0.1111326-0.0695827i$ &\\
      & $\pm0.126106-0.117767i$ & $\pm0.11054-0.11598i$ &  \\
       & $\pm0.124481-0.164874i$ & $\pm0.10964-0.16241i$ &  \\
      \hline
      \end{tabular}
    \caption{Comparison of QNMs in the RN-dS blackhole obtained from the 6pt connection formula with numerical data obtained from the {\it QNMspectral} package.}
    \label{RN6pt:QNMs}
\end{table}

It is completely parallel to also obtain the leading-order analytic form for the PS modes in this region. This corrects the expression given in \eqref{PS-ana} by adding $s$-dependence,
\begin{equation}
    \omega_{\rm PS}=\frac{6(1-9\Lambda)^{\frac{1}{2}}+Q^2(1-9\Lambda)^{-\frac{1}{2}}}{18\sqrt{3}}\lt[\pm\frac{1}{2}\sqrt{4\ell^2+4\ell-1+\frac{14s^2}{3}-\frac{2s^4}{3}}-\lt(n+\frac{1}{2}\rt)i\rt]+\dots,\label{PS-ana-s}
\end{equation}
where again $n=0,1,2,\dots\in\mathbb{N}$. We note that the leading-order analytic expression again gives even better results than the brute-force connection formula computation including higher-order instanton corrections when $\Lambda M^2\approx 0.1$ (see Table \ref{RN6pt:analytic}), which suggests that re-expansion is necessary for the connection formula to write down a series expansion with a clear convergence radius.

\begin{table}
    \centering
    \begin{tabular}{|c|c|c|c|}\hline
      & Conformal block data & QNM package & Leading analytic results \\\hline
      $\Lambda M^2=0.105$ & $\pm0.0815791-0.0235535i$ & $\pm0.0614357-0.0231477i$ & $\pm0.0771136 - 0.0232506i$ \\
      $s=1$,$l=1$ & $\pm0.080743-0.0706604i$ & $\pm0.0609648-0.0694504i$ & $\pm0.0771136 - 0.0697518i$ \\
      & $\pm0.0790442-0.117767i$ & $\pm0.060001-0.115778i$ & $\pm0.0771136 - 0.116253i$\\
      \hline
      $\Lambda M^2=0.105$ & $\pm0.12771-0.0235535i$ & $\pm0.1114268-0.0231930i$ & $\pm0.120814 - 0.0232506i$\\
      $s=1$,$l=2$ & $\pm0.127178-0.0706604i$ & $\pm0.1111326-0.0695827i$ & $\pm0.120814 - 0.0697518i$ \\
      & $\pm0.126106-0.117767i$ & $\pm0.11054-0.11598i$ & $\pm0.120814 - 0.116253i$\\
      \hline
      \end{tabular}
    \caption{Comparison of QNMs in the RN-dS blackhole obtained from the 6pt connection formula with the leading-order analytic results. We see that although only the leading contribution is taken in the analytic expression, it still gives slightly better numerical results than the brute-force instanton computation.}
    \label{RN6pt:analytic}
\end{table}

\section{Conclusion and Discussions}

In this article, we generalized the prescription proposed in \cite{Bonelli:2022ten} to write down the connection formulae for ordinary differential equations with five or six regular singularities. We checked the effectiveness of the connection formulae by computing the QNMs in RN-dS blackholes in both numerical and analytic (series expansion) ways. The analytic expressions of different families of QNMs can be respectively obtained from different coordinate transformations and dictionaries, based on expansion over different parameters. The leading order behaviors of the dS modes and near-extremal modes reproduce the known result conjectured in the literature, and we also provide the analytic expression for PS modes. Our method can systematically be extended to arbitrary high orders. In general, we also saw that even the leading order analytic form of QNMs is supreme to the brute-force computation of connection formula. This suggests that we need to re-organize or re-expand the connection formula, which was originally based on the instanton expansion in quiver gauge theories, in blackhole parameters. In the Heun equation case, similar techniques have been applied in \cite{Lisovyy:2022flm}. We wish to provide such a perturbative connection formula and apply it to study the strong cosmic censorship in a purely analytic way in our future work (including the instability of the fundamental QNM in dS blackholes in various dimensions \cite{Rahman:2018oso,Rahman:2020guv,Sarkar:2023rhp}). It will also help to analyze the connection formula in the colliding limit (c.f. a program initiated in \cite{Zhao:2024rrw}), where irregular singularities can further be put into the differential equation by colliding several regular singularities at a coincident point. We also remark that one can go beyond to consider the dS-Kerr-Newman blackhole with both charge and rotation parameters. The radial part of the Teukolsky equation can also be mapped to an ODE with five singularities \cite{Suzuki:1998vy} (see Appendix \ref{a:dS-Kerr} for the dictionary). Due to the complication of the angular and radial equations coupled together, we will leave it to future work to investigate the numerical and analytic results of the QNMs of such complicated blackhole geometries.  

\paragraph{Acknowledgment} We thank Hongfei Shu for helpful discussions. R.Z. is supported by National Natural Science Foundation of China No. 12105198 and the High-level personnel project of Jiangsu Province (JSSCBS20210709).

\appendix

\section{Nekrasov instanton partition functions for quiver SCFTs and AGT relation}\label{a:quiver}

Let us explain how to use the AGT relation proposed in \cite{Alday:2009aq} to compute the $n$-pt conformal block in this Appendix. 

According to the AGT relation, the conformal blocks of 2d Liouville theory can be computed from the instanton partition functions of the corresponding 4d $\cN=2$ (class ${\cal S}$) superconformal field theories with gauge group SU(2). The instanton partition function can be systematically written down using the Nekrasov factor as building blocks \cite{Nekrasov:2002qd,Flume:2002az,Nekrasov:2012xe},  
\begin{align}
    N_{\lambda\nu}(a,\epsilon_1,\epsilon_2):=\prod_{(i,j)\in\lambda}\lt(a+\epsilon_1(-\nu^t_j+i)+\epsilon_2(\lambda_i-j+1)\rt)\cr
    \times \prod_{(i,j)\in\nu}\lt(a+\epsilon_1(\lambda^t_j-i+1)+\epsilon_2(-\nu_i+j)\rt).
\end{align}
The full expression of the instanton partition function can be found by multiplying the contributions from vector multiplets and matter hypermultiplets (in different representations), and then summing over all possible configurations of the instantons labeled by a tuple of Young diagrams. The SU(2) vector multiplet contribution reads 
\begin{equation}
    Z^{\rm vect}_{\mu\nu}(a)=N^{-1}_{\mu\mu}(1,\epsilon_1,\epsilon_2)N^{-1}_{\mu\nu}(2a,\epsilon_1,\epsilon_2)N^{-1}_{\nu\nu}(1,\epsilon_1,\epsilon_2)N^{-1}_{\nu\mu}(-2a,\epsilon_1,\epsilon_2).
\end{equation}
Our computation involves three types of hypermultiplets, respectively in fundamental, anti-fundamental representations of one gauge group, and bi-fundamental representations of two gauge groups (i.e. fundamental in one gauge group and anti-fundamental in the other). The fundamental and anti-fundamental hypermultiplet contributions are respectively given by 
\begin{align}
    &Z^{\rm fund}_{\mu\nu}(a,\{m_f\}):=\prod_{f=1}^{N_f}N_{\mu\emptyset}(a-m_f,\epsilon_1,\epsilon_2)N_{\nu\emptyset}(-a-m_f,\epsilon_1,\epsilon_2),\\
   &Z^{\rm anti}_{\mu\nu}(a,\{m_f\}):=(-1)^{|\mu|+|\nu|}\prod_{f=1}^{N_{f'}}N_{\emptyset\mu}(-a+m_f-4\epsilon_+,\epsilon_1,\epsilon_2)N_{\emptyset\nu}(a+m_f-4\epsilon_+,\epsilon_1,\epsilon_2),
\end{align}
where $\epsilon_+:=\frac{\epsilon_1+\epsilon_2}{2}$. At the practical level, one can realize the anti-fundamental contributions with the fundamental contributions by substituting $2\epsilon_+-m$. 
\begin{equation}
    Z^{\rm anti}_{\mu\nu}(a,\{m_f\})=Z^{\rm fund}_{\mu\nu}(a,\{2\epsilon_+-m_f\}).\label{fund-anti}
\end{equation}
The bifundamental contribution is given by 
\begin{align}
    Z^{\rm bifund}_{(\mu\nu),(\lambda\rho)}(a,b,m_{\rm bifund})=N_{\mu\lambda}(a-b-m_{\rm bifund},\epsilon_1,\epsilon_2)N_{\mu\rho}(a+b-m_{\rm bifund},\epsilon_1,\epsilon_2)\cr
    \times N_{\nu\lambda}(-a-b-m_{\rm bifund},\epsilon_1,\epsilon_2)N_{\nu\rho}(-a+b-m_{\rm bifund},\epsilon_1,\epsilon_2).
\end{align}

Superconformal field theories used in this article is of linear-quiver type, in which one can use a quiver to indicate the field contents of the theory. 
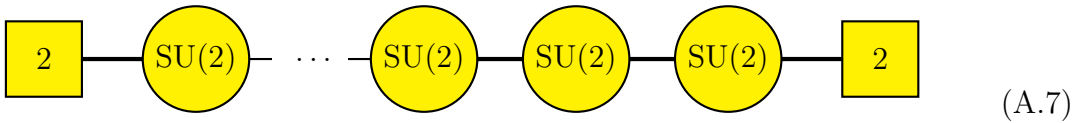
\begin{align}
    \begin{tikzpicture}
    \draw[ultra thick] (0,0)--(2,0);
    \draw[ultra thick] (0,0)--(-2,0);
    \draw[ultra thick] (2,0)--(4,0);
    \draw[ultra thick] (-7,0)--(-5,0);
    \draw[thick] (-2,0)--(-3,0);
    \draw[thick] (-5,0)--(-4,0);
        \draw[fill=yellow,thick] (0,0) circle (0.7);
        \draw[fill=yellow,thick] (2,0) circle (0.7);
        \draw[fill=yellow,thick] (-2,0) circle (0.7);
        \draw[fill=yellow,thick] (-5,0) circle (0.7);
        \draw[fill=yellow,thick] (3.5,0.5) rectangle (4.5,-0.5);
        \draw[fill=yellow,thick] (-7.5,0.5) rectangle (-6.5,-0.5);
        \node at (-2,0) {SU(2)};
        \node at (0,0) {SU(2)};
        \node at (2,0) {SU(2)};
        \node at (-5,0) {SU(2)};
        \node at (4,0) {2};
        \node at (-7,0) {2};
        \node at (-3,0) [left] {\dots};
    \end{tikzpicture}
    \label{linear-quiver}
\end{align}
In the above quiver diagram, each circle denotes a gauge node that represents a vector multiplet containing an SU(2) gauge field, and each thick line represents a hypermultiplet. Lines connecting two gauge nodes correspond to bifundamental hypermultiplets, and squares denote the global flavor symmetry with lines connecting one of the squares giving hypermutiplets in fundamental/anti-fundamental representations. To compute the $n$-pt conformal block, we need to consider a quiver gauge theory with $n-3$ gauge nodes.

When $n=4$, the corresponding theory is called SU(2) gauge theory with $N_f=4$ flavors, and its instanton partition function is given by 
\begin{equation}
    Z^{\rm inst}_{SU(2)\ N_f=4}(a,\{m_f\}_{f=1}^4;\mathfrak{q})=\sum_{\mu,\nu}\mathfrak{q}^{|\mu|+|\nu|}Z^{\rm vect}_{\mu\nu}(a)Z^{\rm fund}_{\mu\nu}(a,\{m_f\}_{f=1}^4),
\end{equation}
where we used \eqref{fund-anti} to convert the anti-fundamental hypermultiplet contributions to fundamental ones by flipping the anti-fudamental mass parameters to $m\rightarrow 2\epsilon_+-m$. For $n>4$, the gauge theory becomes a quiver gauge theory, and the explicit expressions of the instanton partition functions for $n=5$ and $n=5$ are respectively given by 
\begin{align}
    Z^{\rm inst}_{n=5}(a,b,\{m_f\}_{f=1}^4,m_{\rm bifund};\mathfrak{q}_1,\mathfrak{q}_2)=\sum_{\mu,\nu,\lambda,\rho}\mathfrak{q}_1^{|\mu|+|\nu|}\mathfrak{q}_2^{|\lambda|+|\rho|}Z^{\rm fund}_{\mu\nu}(a,\{m_f\}_{f=1}^2)Z^{\rm vect}_{\mu\nu}(a)\cr
    \times Z^{\rm bifund}_{(\mu\nu),(\lambda,\rho)}(a,b,m_{\rm bifund})Z^{\rm vect}_{\lambda\rho}(b)Z^{\rm fund}_{\lambda\rho}(b,\{m_f\}_{f=3}^4),
\end{align}
and 
\begin{align}
    Z^{\rm inst}_{n=6}(a,b,d,\{m_f\}_{f=1}^4,m^{(1)}_{\rm bifund},m^{(2)}_{\rm bifund};\mathfrak{q}_1,\mathfrak{q}_2,\mathfrak{q}_3)=\sum_{\mu,\nu,\lambda,\rho,\tau,\sigma}\mathfrak{q}_1^{|\mu|+|\nu|}\mathfrak{q}_2^{|\lambda|+|\rho|}\mathfrak{q}_3^{|\tau|+|\sigma|}\cr
    \times Z^{\rm fund}_{\mu\nu}(a,\{m_f\}_{f=1}^2)Z^{\rm vect}_{\mu\nu}(a) Z^{\rm bifund}_{(\mu\nu),(\lambda,\rho)}(a,b,m^{(1)}_{\rm bifund})Z^{\rm vect}_{\lambda\rho}(b)\cr
    \times Z^{\rm bifund}_{(\lambda,\rho),(\tau,\sigma)}(b,d,m^{(2)}_{\rm bifund})Z^{\rm vect}_{\tau\sigma}(d)
    Z^{\rm fund}_{\tau\sigma}(d,\{m_f\}_{f=3}^4).
\end{align}
A tricky point to generate the $n$-pt conformal blocks systematically from the instanton partition functions is that it further involves additional U(1) factors. These U(1) factors are given by 
\begin{align}
    F^{n=4}_{\rm U(1)}=&(1-\mathfrak{q})^{(m_1+m_2)\lt(b+b^{-1}-\frac{m_3+m_4}{2}\rt)},\\
    F^{n=5}_{\rm U(1)}=&(1-\mathfrak{q}_1)^{(m_1+m_2)\lt(b+b^{-1}-m_{\rm bifund}\rt)}(1-\mathfrak{q}_2)^{2m_{\rm bifund}\lt(b+b^{-1}-\frac{m_3+m_4}{2}\rt)}\cr
    &\times (1-\mathfrak{q}_1\mathfrak{q}_2)^{(m_1+m_2)\lt(b+b^{-1}-\frac{m_3+m_4}{2}\rt)},\\
    F^{n=6}_{\rm U(1)}=&(1-\mathfrak{q}_1)^{(m_1+m_2)\lt(b+b^{-1}-m^{(1)}_{\rm bifund}\rt)}(1-\mathfrak{q}_2)^{2m^{(1)}_{\rm bifund}\lt(b+b^{-1}-m^{(2)}_{\rm bifund}\rt)}\cr
    & (1-\mathfrak{q}_3)^{2m^{(2)}_{\rm bifund}\lt(b+b^{-1}-\frac{m_3+m_4}{2}\rt)}(1-\mathfrak{q}_1\mathfrak{q}_2)^{(m_1+m_2)\lt(b+b^{-1}-m^{(2)}_{\rm bifund}\rt)}\cr
    & (1-\mathfrak{q}_2\mathfrak{q}_3)^{2m^{(1)}_{\rm bifund}\lt(b+b^{-1}-\frac{m_3+m_4}{2}\rt)}(1-\mathfrak{q}_1\mathfrak{q}_2\mathfrak{q}_3)^{(m_1+m_2)\lt(b+b^{-1}-\frac{m_3+m_4}{2}\rt)}.
\end{align}
The $n$-pt conformal block is proposed from the AGT relation to be given by 
\begin{equation}
    \overline{\mathfrak{F}}=\lt(F^{n}_{\rm U(1)}\rt)^{-1}Z^{\rm inst}_{n},
\end{equation}
where $\overline{\mathfrak{F}}$ is the normalized conformal block starting from $1$ in $\{\mathfrak{q}_i\}$-expansion. For example for $n=5$, from \eqref{confb-5pt} we have 
\begin{align}
    &\mathfrak{F}(\Delta_5,\Delta_4,\Delta_d,\Delta_3,\Delta_a,\Delta_2,\Delta_1;\mathfrak{q}_1,\mathfrak{q}_2)\cr
    &=\mathfrak{q}_1^{\Delta_a-\Delta_2-\Delta_1}\mathfrak{q}_2^{\Delta_d-\Delta_3-\Delta_2-\Delta_1}\overline{\mathfrak{F}}(\Delta_5,\Delta_4,\Delta_d,\Delta_3,\Delta_a,\Delta_2,\Delta_1;\mathfrak{q}_1,\mathfrak{q}_2).
\end{align}

\section{More connection formulae}\label{a:conn}

In this section, we deduce two more connection formulae for the 5-pt BPZ equation, based on the assumption that the following crossing symmetry operation with the connection coefficient ${\cal M}_{\theta\theta'}$ given by \eqref{M-def} can be performed locally in any conformal blocks, 
\begin{align}
\begin{tikzpicture}[x=0.75pt,y=0.75pt,yscale=-0.75,xscale=0.75]
\draw [line width=1.5]  [dash pattern={on 1.69pt off 2.76pt}]  (55,54) -- (113,110) ;
\draw [line width=1.5]    (113,110) -- (57,165) ;
\draw [line width=1.5]    (237,55) -- (181,110) ;
\draw [line width=1.5]    (113,110) -- (181,110) ;
\draw [line width=1.5]    (181,110) -- (239,166) ;
\draw [line width=1.5]  [dash pattern={on 1.69pt off 2.76pt}]  (471,14) -- (529,70) ;
\draw [line width=1.5]    (585,15) -- (529,70) ;
\draw [line width=1.5]    (529,146) -- (473,201) ;
\draw [line width=1.5]    (529,70) -- (529,146) ;
\draw [line width=1.5]    (529,146) -- (587,202) ;
\draw (286,88.4) node [anchor=north west][inner sep=0.75pt]    {$=\ \sum _{\theta '=\pm }\mathcal{M}_{\theta \theta '}( a_{1} ,a_{2} ,a_{3})$};
\draw (34,32.4) node [anchor=north west][inner sep=0.75pt]    {$z$};
\draw (36,168.4) node [anchor=north west][inner sep=0.75pt]    {$1$};
\draw (246,37.4) node [anchor=north west][inner sep=0.75pt]    {$2$};
\draw (247,170.4) node [anchor=north west][inner sep=0.75pt]    {$3$};
\draw (457,200.4) node [anchor=north west][inner sep=0.75pt]    {$1$};
\draw (593,196.4) node [anchor=north west][inner sep=0.75pt]    {$3$};
\draw (454,18.4) node [anchor=north west][inner sep=0.75pt]    {$z$};
\draw (592,24.4) node [anchor=north west][inner sep=0.75pt]    {$2$};
\end{tikzpicture}
\label{crossing}
\end{align}
This gives a further generalization of the trick introduced in \cite{Bonelli:2022ten}. 

A connection formula we wish to derive and is used in the main text is the one connecting between $z\sim 0$ to $z\sim q$. The deduction can be realized and split into two steps shown below:
\begin{align}
\begin{tikzpicture}[x=0.75pt,y=0.75pt,yscale=-1,xscale=1]
\draw [line width=1.5]  [dash pattern={on 1.69pt off 2.76pt}]  (236,74) -- (236,132) ;
\draw [line width=1.5]    (28,131) -- (274,131) ;
\draw [line width=1.5]    (182,73) -- (182,131) ;
\draw [line width=1.5]    (126,73) -- (126,131) ;
\draw [line width=1.5]    (69,73) -- (69,131) ;
\draw [line width=1.5]  [dash pattern={on 1.69pt off 2.76pt}]  (559,77) -- (559,135) ;
\draw [line width=1.5]    (406,133) -- (652,133) ;
\draw [line width=1.5]    (613,75) -- (613,133) ;
\draw [line width=1.5]    (504,75) -- (504,133) ;
\draw [line width=1.5]    (447,75) -- (447,133) ;
\draw (263,82.4) node [anchor=north west][inner sep=0.75pt]    {$=\ \sum _{\theta '=\pm }\mathcal{M}_{\theta \theta '}( a_{0} ,a ,a_{t})$};
\draw (231,53.4) node [anchor=north west][inner sep=0.75pt]    {$z$};
\draw (258,141.4) node [anchor=north west][inner sep=0.75pt]    {$0$};
\draw (178,53.4) node [anchor=north west][inner sep=0.75pt]    {$t$};
\draw (121,51.4) node [anchor=north west][inner sep=0.75pt]    {$q$};
\draw (64,54.4) node [anchor=north west][inner sep=0.75pt]    {$1$};
\draw (33,143.4) node [anchor=north west][inner sep=0.75pt]    {$\infty $};
\draw (632,137.4) node [anchor=north west][inner sep=0.75pt]    {$0$};
\draw (407,139.4) node [anchor=north west][inner sep=0.75pt]    {$\infty $};
\draw (609,54.4) node [anchor=north west][inner sep=0.75pt]    {$t$};
\draw (556,54.4) node [anchor=north west][inner sep=0.75pt]    {$z$};
\draw (499,52.4) node [anchor=north west][inner sep=0.75pt]    {$q$};
\draw (442,55.4) node [anchor=north west][inner sep=0.75pt]    {$1$};
\draw (148,143.4) node [anchor=north west][inner sep=0.75pt]    {$a$};
\draw (92,145.4) node [anchor=north west][inner sep=0.75pt]    {$d$};
\draw (205,143.4) node [anchor=north west][inner sep=0.75pt]    {$a_{0\theta }$};
\draw (523,138.4) node [anchor=north west][inner sep=0.75pt]    {$a$};
\draw (467,140.4) node [anchor=north west][inner sep=0.75pt]    {$d$};
\draw (580,138.4) node [anchor=north west][inner sep=0.75pt]    {$a_{\theta '}$};
\end{tikzpicture}
\label{0-to-mid}
\end{align}
\begin{align}
\begin{tikzpicture}[x=0.75pt,y=0.75pt,yscale=-1,xscale=1]
\draw [line width=1.5]  [dash pattern={on 1.69pt off 2.76pt}]  (177,102) -- (177,160) ;
\draw [line width=1.5]    (24,158) -- (270,158) ;
\draw [line width=1.5]    (231,100) -- (231,158) ;
\draw [line width=1.5]    (122,100) -- (122,158) ;
\draw [line width=1.5]    (65,100) -- (65,158) ;
\draw [line width=1.5]  [dash pattern={on 1.69pt off 2.76pt}]  (590,77) -- (529,77) ;
\draw [line width=1.5]    (406,158) -- (652,158) ;
\draw [line width=1.5]    (613,100) -- (613,158) ;
\draw [line width=1.5]    (529,42) -- (529,158) ;
\draw [line width=1.5]    (447,100) -- (447,158) ;
\draw (261,108.4) node [anchor=north west][inner sep=0.75pt]    {$=\ \sum _{\theta '=\pm }\mathcal{M}_{( -\theta ) \theta '}( a ,a_{q} ,d)$};
\draw (250,162.4) node [anchor=north west][inner sep=0.75pt]    {$0$};
\draw (25,164.4) node [anchor=north west][inner sep=0.75pt]    {$\infty $};
\draw (227,79.4) node [anchor=north west][inner sep=0.75pt]    {$t$};
\draw (174,79.4) node [anchor=north west][inner sep=0.75pt]    {$z$};
\draw (117,77.4) node [anchor=north west][inner sep=0.75pt]    {$q$};
\draw (60,80.4) node [anchor=north west][inner sep=0.75pt]    {$1$};
\draw (141,163.4) node [anchor=north west][inner sep=0.75pt]    {$a_{-\theta }$};
\draw (85,165.4) node [anchor=north west][inner sep=0.75pt]    {$d$};
\draw (198,163.4) node [anchor=north west][inner sep=0.75pt]    {$a$};
\draw (632,162.4) node [anchor=north west][inner sep=0.75pt]    {$0$};
\draw (407,164.4) node [anchor=north west][inner sep=0.75pt]    {$\infty $};
\draw (609,79.4) node [anchor=north west][inner sep=0.75pt]    {$t$};
\draw (565,53.4) node [anchor=north west][inner sep=0.75pt]    {$z$};
\draw (505,39.4) node [anchor=north west][inner sep=0.75pt]    {$q$};
\draw (442,80.4) node [anchor=north west][inner sep=0.75pt]    {$1$};
\draw (563,163.4) node [anchor=north west][inner sep=0.75pt]    {$a$};
\draw (484,166.4) node [anchor=north west][inner sep=0.75pt]    {$d$};
\draw (493,103.4) node [anchor=north west][inner sep=0.75pt]    {$a_{q\theta '}$};
\end{tikzpicture}
\label{mid-to-q}
\end{align}
In the first step \eqref{0-to-mid}, the original conformal block $\mathfrak{F}(\Delta_5,\Delta_4,\Delta_d,\Delta_3,\Delta_a,\Delta_2,\Delta_{1\theta},\Delta_{2,1},\Delta_1;z/t,t/q,q)$ is decomposed into the linear combination of 
\begin{equation}
    \mathfrak{F}(\Delta_5,\Delta_4,\Delta_d,\Delta_3,\Delta_a,\Delta_{2,1},\Delta_{a\theta'},\Delta_2,\Delta_1;t/z,z/q,q)
\end{equation}
The second step further involves a coordinate transformation, 
\begin{equation}
    z\to \frac{z-q}{1-q},
\end{equation}
so that the conformal blocks 
\begin{equation}
    \mathfrak{F}\lt(\Delta_5,\Delta_4,\Delta_d,\Delta_{3\theta'},\Delta_3,\Delta_{2,1},\Delta_a,\Delta_2,\Delta_1;\frac{z-q}{t-q},\frac{t-q}{q},\frac{q}{q-1}\rt),
\end{equation}
will be given as expansions around $z\sim q$. In the semiclassical limit $b\to 0$, one obtains the following connection formula, 
\begin{align}
    {\cal F}_\theta(t/z,z/q,q)=\sum_{\theta'=\pm}\lt(\sum_{\sigma}\cM_{\theta\sigma}(a_0,a,a_t)\cM_{(-\sigma)\theta'}(a,a_q,d)\lt(\frac{t}{1-q}\rt)^{-\sigma a}e^{-\frac{\sigma}{2}\partial_a F}\rt)\cr
    \times (q-1)^{\frac{1}{2}}{\cal F}_{\theta'}\lt(\frac{z-q}{t-q},-\frac{t-q}{q},\frac{q}{q-1}\rt).\label{eq:conn-0-q}
\end{align}
It is also straightforward to write down a connection formula between $z\sim 0$ and $z\sim 1$ in a similar manner or between $z\sim 0$ and $z\sim \infty$ (by using the crossing relation \eqref{crossing} three times). Since we do not need them in our work, we will not write down the explicit form of these connection formulae.  



\section{dS-Kerr-Newman blackhole and 5-pt BPZ equation}\label{a:dS-Kerr}

In this Appendix, we show that the perturbation equations of the dS-Kerr-Newman blackhole can be brought to ODEs with 4 and 5 regular singularities. 

In the Boyer-Lindquist coordinates the Kerr-Newman de-Sitter metric takes the form \cite{Carter:1968ks,Suzuki:1998vy},
\begin{align}
    ds^2=-\rho^2\lt(\frac{{\rm d}r^2}{\Delta_r}+\frac{{\rm d}\theta^2}{\Delta_\theta}\rt)-\frac{\Delta_\theta \sin^2\theta}{(1+\alpha)^2\rho^2}\lt[a{\rm d}t-(r^2+a^2){\rm d}\varphi\rt]^2
    +\frac{\Delta_r}{(1+\alpha)^2\rho^2}({\rm d}t-a\sin^2\theta d\varphi)^2\label{KN:metric}
\end{align}
where $a$ is the rotation parameter, and 
\begin{align}
    \Delta_r=(r^2+a^2)(1-\frac{\alpha}{a^2}r^2)-2Mr+Q^2=-\frac{\alpha}{a^2}(r-r_-)(r-r_0)(r-r_+)(r-r_c),\cr
    \Delta_\theta=1+\alpha \cos^2\theta,\quad\alpha=\frac{\Lambda a^2}{3},\quad\bar{\rho}=r+ia\cos\theta,\quad\rho^2=\bar{\rho}\bar{\rho}^*.
\end{align}
After separating the Teukolsky equation into the angular part and the radial part, we obtain the angular equation as 
\begin{align}
    \lt(\frac{\rm d}{{\rm d}x}(1+\alpha x^2)(1-x^2)\frac{\rm d}{{\rm d}x}+\lambda-s(1-\alpha)+\frac{(1+\alpha)^2}{\alpha}\xi^2-2\alpha x^2\rt.\cr
    +\frac{1+\alpha}{1+\alpha x^2}[2s(\alpha m-(1+\alpha)\xi)x-\frac{(1+\alpha)^2}{\alpha}\xi^2-2m(1+\alpha)\xi+s^2]\cr
    \lt.-\frac{(1+\alpha)^2m^2}{(1+\alpha x^2)(1-x^2)}-\frac{(1+\alpha)(s^2+2smx)}{1-x^2}\rt)S(x)=0\label{KNeq:Angular}
\end{align}
where $x=\cos\theta$, $\xi=a\omega$, $m$ is the angular momentum in the $z$-direction, $\lambda$ corresponds to the angular momentum in the $a\to 0$ limit, and the radial part is given by 
\begin{align}
    \lt(\Delta_r^{-s}\frac{\rm d}{{\rm d}r}\Delta_r^{s+1}\frac{\rm d}{{\rm d}r}+\frac{1}{\Delta_r}\lt[(1+\alpha)^2\lt(K-\frac{eQr}{1+\alpha}\rt)^2-is(1+\alpha)\lt(K-\frac{eQr}{1+\alpha}\rt)\frac{{\rm d}\Delta_r}{{\rm d}r}\rt]\rt.\cr
    \lt.+4is(1+\alpha)\omega r-\frac{2\alpha}{a^2}(s+1)(2s+1)r^2+2s(1-\alpha)-2iseQ-\lambda\rt)R(r)=0,
\end{align}
with $K=\omega(r^2+a^2)-am$ and $e$ representing the charge of the massless field considered.  

\paragraph{Dictionary for angular part}
The equation \eqref{KNeq:Angular} has four singularities at -1,1,$-\frac{i}{\sqrt{\alpha}}$ and $\frac{i}{\sqrt{\alpha}}$. Taking the coordinate transform, 
\begin{align}
    z=\frac{(-\frac{i}{\sqrt{\alpha}}-\frac{i}{\sqrt{\alpha}})(x+1)}{(-\frac{i}{\sqrt{\alpha}}+1)(x-\frac{i}{\sqrt{\alpha}})},
\end{align}
one can map $x=-1$ to $z=0$, $x=-\frac{i}{\sqrt{\alpha}}$ to $z=1$, $x=\frac{i}{\sqrt{\alpha}}$ to $z=\infty$ and $x=1$ to $z=t$,
where
\begin{align}
    t=-\frac{4i\sqrt{\alpha}}{(-i+\sqrt{\alpha})^2}.
\end{align}
Bring into the Sch\"odinger form and we obtain the dictionary of 4-pt BPZ equation with the potential,  
\begin{align}
    Q_{SW}(z)=\frac{\alpha_4}{z^2}+\frac{\beta_4}{(z-t)^2}+\frac{\gamma_4}{(z-1)^2}+\frac{\delta_4}{z(z-1)}
    +\frac{\eta_4}{z(z-t)},
\end{align}
where 
\begin{align}
    \alpha_4=\frac{1}{4}(1-m^2+2ms-s^2),\quad\beta_4=-\frac{1}{4}(-1+m+s)(1+m+s),
\end{align}
\begin{align}
    \gamma_4=\frac{\alpha-s^2\alpha+2is\alpha^\frac{3}{2}(m-\xi)-2is\sqrt{\alpha}\xi+\xi^2+2\alpha\xi(m+\xi)+\alpha^2(m+\xi)^2}{4\alpha},
\end{align}
\begin{align}
    \delta_4=\frac{\sqrt{\alpha}(s+s^2(-1+\alpha)+2m^2\alpha-s\alpha-ims\sqrt{\alpha}(1+\alpha)-\lambda)+(is(-1+\alpha)+2m\sqrt{\alpha})(1+\alpha)\xi}{(i+\sqrt{\alpha})^2\sqrt{\alpha}},
\end{align}
\begin{align}
    \eta_4=-(s^2(-i+\sqrt{\alpha})^2+(i+\sqrt{\alpha})^2+m^2(1-2i\sqrt{\alpha}+3\alpha)-2\lambda+4m(-s(i\sqrt{\alpha}+\alpha)\cr
    +\xi+\alpha\xi)+s(2-2\alpha+4\xi+4\alpha\xi))/(2(i+\sqrt{\alpha})^2).
\end{align}

\paragraph{Dictionary for radial part}
Similar to the RN-dS blackhole, one can set
\begin{align}
    z=\frac{(r_0-r_-)(r-r_+)}{(r-r_-)(r_0-r_+)}
\end{align}
to map $r=r_-$ to the infinity, $r=r_0$ to $z=1$, $r=r_+$ to $z=0$, $r=r_c$ mapped to $z=t$ with 
\begin{align}
    t=\frac{(r_0-r_-)(r_c-r_+)}{(r_c-r_-)(r_0-r_+)}
\end{align}
and one more regular singularity at $z=q$ with 
\begin{align}
    q=\frac{r_0-r_-}{r_0-r_+}.
\end{align}
Bring the radial equation into the Sch\"odinger form, one obtains the following dictionary,
\begin{align}
    Q_{SW}(z)=\frac{\alpha_5}{z^2}+\frac{\beta_5}{(z-t)^2}+\frac{\gamma_5}{(z-q)^2}+\frac{\delta_5}{(z-1)^2}+\frac{\eta_5}{z(z-1)}\cr
    +\frac{\kappa_5}{z(z-t)}+\frac{\mu_5}{z(z-q)},
\end{align}
with 
\begin{align}
    \alpha_5=(-(r_0-r_+)^2(r_c-r_+)^2(r_--r_+)^2(-1+s^2)\alpha^2-2ia^3m(r_0-r_+)(-r_c+r_+)\cr
    (-r_-+r_+)(-2+s)s\alpha(1+\alpha)-2a^7m(4+s^2)(1+\alpha)^2\omega+a^8(4+s^2)(1+\alpha)^2\omega^2\cr
    +2ia^2(r_0-r_+)r_+(-r_c+r_+)(-r_-+r_+)(-2+s)s\alpha(-eQ+r_+(1+\alpha)\omega)-\cr
    2a^5mr_+(4+s^2)(1+\alpha)(-eQ+r_+(1+\alpha)\omega)+a^6(4+s^2)(1+\alpha)(m^2(1+\alpha)+\cr
    2r_+\omega(-eQ+r_+(1+\alpha)\omega))+a^4(e^2Q^2r_+^2(4+s^2)-2eQr_+^3(4+s^2)(1+\alpha)\omega\cr
    +(1+\alpha)\omega(2i(r_0-r_+)(-r_c+r_+)(-r_-+r_+)(-2+s)s\alpha+r_+^4(4+s^2)(1+\alpha)\omega)))\cr
    /(4(r_0-r_+)^2(r_c-r_+)^2(r_--r_+)^2\alpha^2),
\end{align}
\begin{align}
    \beta_5=(-(r_0-r_c)^2(r_c-r_-)^2(r_c-r_+)^2(-1+s^2)\alpha^2-2ia^3m(r_0-r_c)(r_c-r_-)\cr
    (r_c-r_+)(-2+s)s\alpha(1+\alpha)-2a^7m(4+s^2)(1+\alpha)^2\omega+a^8(4+s^2)(1+\alpha)^2\omega^2\cr
    +2ia^2(r_0-r_c)r_c(r_c-r_-)(r_c-r_+)(-2+s)s\alpha(-eQ+r_c(1+\alpha)\omega)-\cr
    2a^5mr_c(4+s^2)(1+\alpha)(-eQ+r_c(1+\alpha)\omega)+a^6(4+s^2)(1+\alpha)(m^2(1+\alpha)+\cr
    2r_c\omega(-eQ+r_c(1+\alpha)\omega))+a^4(e^2Q^2r_c^2(4+s^2)-2eQr_c^3(4+s^2)(1+\alpha)\omega\cr
    +i(1+\alpha)\omega(2r_0(r_c-r_-)(r_c-r_+)(-2+s)s\alpha+r_c(-2r_c^2(-2+s)s\alpha\cr
    -r_-r_+(-2+s)s\alpha+2r_c(r_-+r_+)(-2+s)s\alpha-ir_c^3(4+s^2)(1+\alpha)\omega))))\cr
    /(4(r_0-r_c)^2(r_c-r_-)^2(r_c-r_+)^2\alpha^2),
\end{align}
\begin{align}
    \gamma_5=0,\quad\mu_5=\frac{(1+2s)((r_0+r_c+r_-+r_+)(1+s)\alpha-ia^2s(1+\alpha)\omega)}{(r_--r_+)\alpha},
\end{align}
\begin{align}
    \delta_5=(-(r_0-r_c)^2(r_0-r_-)^2(r_0-r_+)^2(-1+s^2)\alpha^2+2ia^3m(r_0-r_c)(r_0-r_-)\cr
    (r_0-r_+)(-2+s)s\alpha(1+\alpha)-2a^7m(4+s^2)(1+\alpha)^2\omega+a^8(4+s^2)(1+\alpha)^2\omega^2\cr
    -2ia^2r_0(r_0-r_c)(r_0-r_-)(r_0-r_+)(-2+s)s\alpha(-eQ+r_0(1+\alpha)\omega)-\cr
    2a^5mr_0(4+s^2)(1+\alpha)(-eQ+r_0(1+\alpha)\omega)+a^6(4+s^2)(1+\alpha)(m^2(1+\alpha)+\cr
    2r_0\omega(-eQ+r_0(1+\alpha)\omega))+a^4(e^2Q^2r_0^2(4+s^2)-2eQr_0^3(4+s^2)(1+\alpha)\omega\cr
    +(1+\alpha)\omega(-2i(r_0-r_c)(r_0-r_-)(r_0-r_+)(-2+s)s\alpha+r_0^4(4+s^2)(1+\alpha)\omega)))\cr
    /(4(r_0-r_c)^2(r_0-r_-)^2(r_0-r_+)^2\alpha^2).
\end{align}
\begin{align}
    \eta_5=(-(r_0-r_c)^2(r_0-r_-)(r_0-r_+)^2(1+s)(-r_-r_+(1+s)+r_0^2(2+4s)\cr
    -r_c(r_-+r_-s+2r_+s)+r_0(r_c+r_++3r_cs+3r_+s+2m(1+s)))\alpha^2-\cr
    2ia^3m(r_0-r_c)^2(r_0-r_+)^2(-2+s)s\alpha(1+\alpha)-2a^7m(2r_0-r_c-r_+)\cr
    (4+s^2)(1+\alpha)^2\omega+a^8(2r_0-r_c-r_+)(4+s^2)(1+\alpha)^2\omega^2-a^5m\cr
    (4+s^2)(1+\alpha)^2(eQ(-3r_0^2+r_cr_++r_0(r_c+r_+))+2r_0(r_0^2-r_cr_+)\cr
    (1+\alpha)\omega)+a^6(4+s^2)(1+\alpha)(m^2(2r_0-r_c-r_+)(1+\alpha)+\omega(eQ\cr
    (-3r_0^2+r_cr_++r_0(r_c+r_+))+2r_0(r_0^2-r_cr_+)(1+\alpha)\omega))+ia^2(r_0-r_c)^2\cr
    (r_0-r_+)^2\alpha(eQr_-s(3+s)-4ir_-s(-1+\alpha)+2ir_0\lambda-2ir_-\lambda+2r_0^2s\cr
    (1+2s)(1+\alpha)\omega+r_0s(e(Q-3Qs)+4i(-1+\alpha)-2r_-(3+s)\cr
    (1+\alpha)\omega))+a^4(e^2Q^2r_0(r_0^2-r_cr_+)(4+s^2)-eQr_0^2(r_0^2-3r_cr_+\cr
    +r_0(r_c+r_+))(4+s^2)(1+\alpha)\omega)))/(2(r_0-r_c)^3(r_0-r_-)(r_0-r_+)^2\cr
    (r_--r_+)\alpha^2).
\end{align}
\begin{align}
    \kappa_5=((r_0-r_c)^2(r_c-r_-)(r_c-r_+)^2(1+s)(-r_-r_+(1+s)+r_c^2(2+4s)\cr
    +r_0(r_c+3r_cs-2r_+s-r_m(1+s))+r_c(r_++3r_+s+2r_-(1+s)))\alpha^2+\cr
    2ia^3m(r_0-r_c)^2(r_c-r_+)^2(-2+s)s\alpha(1+\alpha)-2a^7m(r_0-2r_c+r_+)\cr
    (4+s^2)(1+\alpha)^2\omega+a^8(r_0-2r_c+r_+)(4+s^2)(1+\alpha)^2\omega^2+a^5m\cr
    (4+s^2)(1+\alpha)^2(eQ(r_c(-3r_c+r_+)+r_0(r_c+r_+))+2r_c(r_c^2-
    r_0r_+)\cr
    (1+\alpha)\omega)+a^6(4+s^2)(1+\alpha)(m^2(r_0-2r_c+r_+)(1+\alpha)+\omega(-eQ\cr
    (r_c(-3r_c+r_+)+r_0(r_c+r_+))-2r_c(r_c^2-r_0r_+)(1+\alpha)\omega))\cr
    -ia^2(r_0-r_c)^2(r_c-r_+)^2\alpha(r_-(3eQs+eQs^2-4is(-1+\alpha)\cr
    -2i\lambda)+2r_c^2s(1+2s)(1+\alpha)\omega-r_c(-2i\lambda+s(4i-eQ-4i\alpha+6r_-\omega+6r_-\cr
    \alpha\omega)+s^2(3eQ+2r_-(1+\alpha)\omega)))+a^4(-e^2Q^2r_c(r_c^2-r_0r_+)(4+s^2)\cr
    +eQr_c^2(r_0(r_c-3r_+)+r_c(r_c+r_+))(4+s^2)(1+\alpha)\omega-i(1+\alpha)\omega\cr
    (2r_0^2(r_c-r_+)^2(-2+s)s\alpha+r_0r_c(8r_cr_+(-2+s)s\alpha-4r_+^2(-2+s)\cr
    s\alpha-ir_c^3(4+s^2)(1+\alpha)\omega+r_c^2(8s\alpha+8ir_+(1+\alpha)\omega+2is^2(2i\alpha+r_+\omega\cr
    +r_+\alpha\omega)))+r_c^2(-4r_cr_+(-2+s)s\alpha+2r_+^2(-2+s)s\alpha-ir_c^2(-4is\alpha\cr
    +4r_+(1+\alpha)\omega+s^2(2i\alpha+r_+\omega+r_+\alpha\omega))))))/(2(r_0-r_c)^3(r_c-r_-)\cr
    (rc-r_+)^2(r_--r_+)\alpha^2).
\end{align}

\bibliographystyle{JHEP}
\bibliography{quiver}

\end{document}